\documentclass[entropy,article,accept,pdftex,moreauthors]{Definitions/mdpi} 

\usepackage[normalem]{ulem}

\firstpage{1} 
\pubvolume{8}
\issuenum{1}
\articlenumber{153}
\pubyear{2022}
\copyrightyear{2022}
\datereceived{31 January 2022} 
\dateaccepted{22 February 2022} 
\datepublished{28 February 2022} 
\hreflink{https://doi.org/10.3390/universe8030153} 

\Title{Compact Objects in Alternative Gravities}

\TitleCitation{Compact Objects}

\Author{Jose Luis Bl\'azquez-Salcedo $^{1}$\orcidA{0000-0002-8156-8372}, Burkhard Kleihaus $^{2}$\orcidA{0000-0001-9751-4400} and Jutta Kunz $^{3}$*\orcidA{0000-0001-7990-8713}}

\AuthorNames{Firstname Lastname, Firstname Lastname and Firstname Lastname}

\AuthorCitation{Blázquez-Salcedo, J.L.;
	Kleihaus, B.; Kunz, J.}

\address{%
$^{1}$ \quad Departamento de F\'isica Te\'orica II and IPARCOS, Facultad de Ciencias F\'isicas, 
Universidad Complutense de Madrid, Spain; jlblaz01@ucm.es\\
$^{2}$ \quad Institute of Physics, University Oldenburg, Germany; b.kleihaus@uni-oldenburg.de\\
$^{3}$ \quad Institute of Physics, University Oldenburg, Germany; jutta.kunz@uni-oldenburg.de
}

\corres{Correspondence: jutta.kunz@uni-oldenburg.de}

\abstract{
We address neutron stars and black holes in alternative gravities, after recalling their basic properties in General Relativity.
Among the plethora of interesting alternative gravities we here focus on an interesting set of scalar-tensor theories.
We discuss the phenomenon of spontaneous scalarization, that is matter induced for neutron stars and curvature induced for black holes.
Along with other relevant physical properties, we address the quasi-normal modes of these compact objects.
In particular, we consider \textit{universal relations} of neutron stars to largely reduce the dependence on the equation of state, and we briefly address the shadow of black holes. 
}

\begin{document}

\section{Introduction}

Alternative gravities have moved into the focus since many years for a variety of reasons, as discussed in numerous contributions at this conference, where cosmological considerations are the focal point.
The present contribution is meant as a supplement to these considerations, putting emphasis on compact objects in alternative gravities (see e.g.~\cite{Faraoni:2010pgm,Berti:2015itd,CANTATA:2021ktz}).
With the advent of gravitational wave observations by LIGO/VIRGO, in combination with multi-messenger astronomy \cite{LIGOScientific:2016aoc,LIGOScientific:2017vwq,LIGOScientific:2017ync} and first observations by the EHT consortium \cite{EventHorizonTelescope:2019dse}, the study of black holes and neutron stars has entered a new golden age of gravitational physics.
The new data and the enhanced future data from upgraded and new observatories will allow to put (stronger) constraints on numerous alternative gravities, that will complement the available solar system data and the data from pulsar observations (see e.g.~\cite{Will:2018bme,Shao:2014wja,Kramer:2021jcw}).

In this contribution we will first address neutron stars and then black holes. 
In both cases we will set the stage by recalling properties of neutron stars and black holes in General Relativity (GR).
Subsequently we will discuss corresponding properties for a selection of alternative gravities, where we focus on metric theories with an additional scalar degree of freedom.
In the case of neutron stars we will put emphasis on \textit{universal relations}, that allow to minimize the effects of the persisting uncertainty of the equation of state of nuclear matter, like the $I$-Love-$Q$ relations and \textit{universal relations} for quasi-normal modes (QNMs)
(see e.g.~\cite{Yagi:2016bkt,Doneva:2017jop}).
The theories we will consider include scalar-tensor theories (STTs) \cite{Brans:1961sx,Damour:1992we,Fujii:2003pa} with spontaneous scalarization \cite{Damour:1993hw},  $R^2$ gravity \cite{Sotiriou:2008rp,DeFelice:2010aj,Nojiri:2010wj,Capozziello:2011et,Yazadjiev:2014cza}, or Chameleon screening \cite{Brax:2012gr}.
For black holes we will address besides their basic properties their shadow and also their QNMs.
We will discuss scalarization in Einstein-scalar-Gauss-Bonnet (EsGB) theories \cite{Zwiebach:1985uq,Gross:1986mw,Metsaev:1987zx}, where the scalar could be a dilaton or some other field, depending on the coupling function.
In the latter case curvature-induced spontaneously scalarized black holes can arise \cite{Doneva:2017bvd,Silva:2017uqg,Antoniou:2017acq}, where for rapid rotation also spin-induced scalarization can be present \cite{Dima:2020yac}.

\section{Neutron Stars}

\subsection{Neutron Stars in General Relativity}

\begin{figure}[ht]
\begin{center}
\mbox{
\includegraphics[width=.65\textwidth, angle =270]{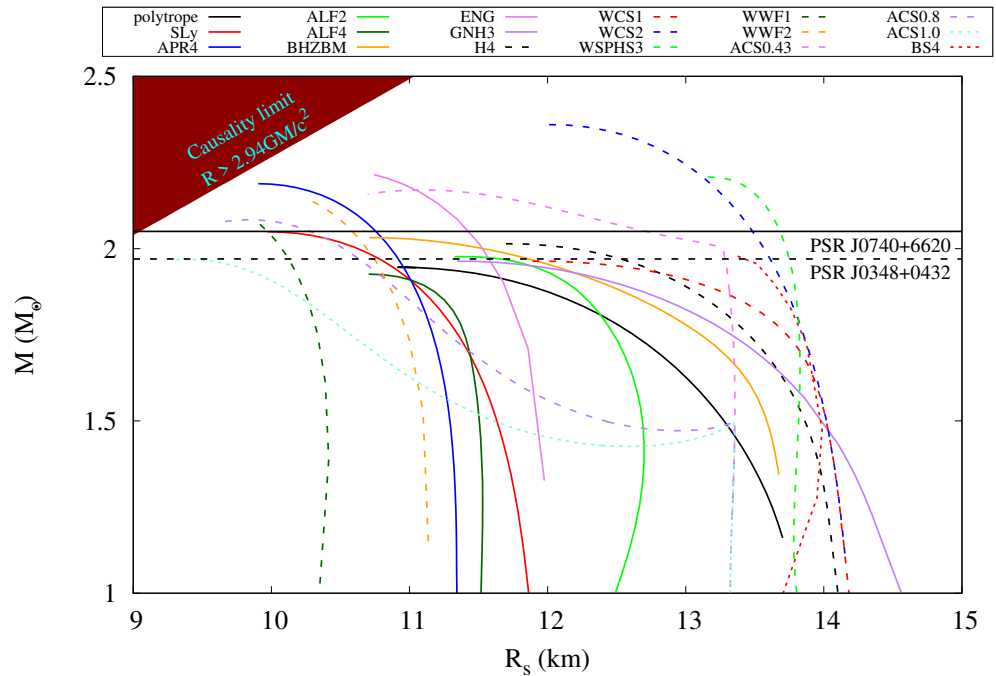}
}
\end{center}
\vspace{-0.5cm}
\caption{{\it{Mass-radius relation: neutron star mass $M$ (in solar masses $M_\odot$) vs radius $R$ (in km) for various equations of state. 
Also shown are high mass pulsars
\cite{Demorest:2010bx,Antoniadis:2013pzd,NANOGrav:2019jur}
and the causality limit.}}}
\label{fig1}
\end{figure}

Neutron stars are highly compact objects as demonstrated in Fig.~\ref{fig1}, where their mass--radius relation is shown for various equations of state. 
Despite strong efforts both from the observational and the theoretical side, there are still large uncertainties concerning the proper description of nuclear matter under the extreme conditions of strong gravity as found in neutron stars (see, e.g., the reviews \cite{Lattimer:2012nd,Ozel:2016oaf,Baym:2017whm}),
although the observations of pulsars with masses of about 2 solar masses have already ruled out numerous equations of state \cite{Demorest:2010bx,Antoniadis:2013pzd,NANOGrav:2019jur}.

The strong dependence of neutron star properties on the equation of state may seem a serious obstacle to employ neutron star data to extract constraints for many of the alternative gravities.
However, here the \textit{universal relations} can come to the rescue \cite{Yagi:2016bkt,Doneva:2017jop}.
These represent relations between neutron star properties, that are to a large extent
independent of the neutron star’s internal structure.
The most impressive relations are those between the moment of inertia $I$, the Love number, i.e., the tidal deformability $\lambda$, and the quadrupole moment $Q$ \cite{Yagi:2013bca}.

Like the mass and the radius, also the moment of inertia, the tidal deformability and the quadrupole moment show a strong dependence on the equation of state (see e.g.~\cite{Krastev:2016wib}).
However, when instead the dimensionless quantities $\bar I=I/M^3$, $\bar \lambda=\lambda/M^5$ and $\bar Q=QM/J^2$ (with spin $J$) are considered, the dependence on the equation of state almost disappears \cite{Yagi:2013bca,Yagi:2016bkt}.
The dimensionless quantities satisfy simple relations of the type 
\begin{equation}
\ln y_i 
= a_i + b_i \ln x_i + c_i (\ln x_i)^2 + d_i (\ln x_i)^3+ e_i (\ln x_i)^4 ,
\label{ILQ}
\end{equation}
the so-called $I$-Love-$Q$ relations, that are satisfied to 1\% or better \cite{Yagi:2013bca,Yagi:2016bkt}.
Moreover, besides the lowest multipole moments also the higher multipole moments satisfy such relations, although the deviations increase somewhat \cite{Yagi:2013bca,Yagi:2016bkt}.
{(See for instance \cite{Manko:2019bgn,Jiang:2019vmf} for some analytical results on such \textit{universal relations}.)}

Whereas the multipole moments of the neutron stars represent important characteristics of the stationary solutions and their tidal reaction to an external companion, further relevant characteristics of neutron stars are obtained from asteroseismology, i.e., the study of their spectra.
The neutron star matter possesses a large variety of modes, that may be excited by external interactions.
Of foremost importance are of course gravitational waves, that will be emitted in the collision of neutrons stars.
In such processes the inspiral phase is followed by the merger and the ringdown, providing ample data for crucial analysis \cite{LIGOScientific:2017ync}.

The spectrum of neutron stars can be investigated by calculating their QNMs (see e.g.~\cite{Andersson:1997rn,Kokkotas:1999bd,Berti:2015itd}),
that should be discernible by future gravitational wave detectors in the ringdown phase after collisions.
Theoretically QNMs are obtained by employing perturbation theory. 
So far, most studies of QNMs of neutron stars have considered static spherically symmetric background solutions, making use of the simplifications arising due to symmetry.
In particular, the perturbation equations then decouple for modes with odd (axial) and even (polar) parity, and they depend only on the angular number $l$ (and not the azimuthal number $m$).

The perturbations concern the metric
\begin{equation}
    g_{\mu\nu} = g_{\mu\nu}^{(0)}(r) + \epsilon h_{\mu\nu}(t,r,\theta,\varphi) \ ,
\end{equation}
and the nuclear matter
\begin{eqnarray}
p &=& p_0(r) + \epsilon \delta p(t,r,\theta,\varphi) \ , \\
\rho &=& \rho_0(r) + \epsilon \delta \rho(t,r,\theta,\varphi)  \\
u &=& u^{(0)} + \epsilon \delta u(t,r,\theta,\varphi) \ ,
\end{eqnarray}
with pressure $p$, density $\rho$ and velocity $u$, and the index $(0)$ denotes the background values.
The resulting system of perturbation equations then corresponds to an eigenvalue problem with a complex eigenvalue $\omega$, whose real part $\omega_R$ and imaginary part $\omega_I$ describe the frequency and the decay rate of the gravitational wave, respectively.

\begin{figure}[ht]
\begin{center}
\mbox{
\includegraphics[width=5.0cm, angle=270]{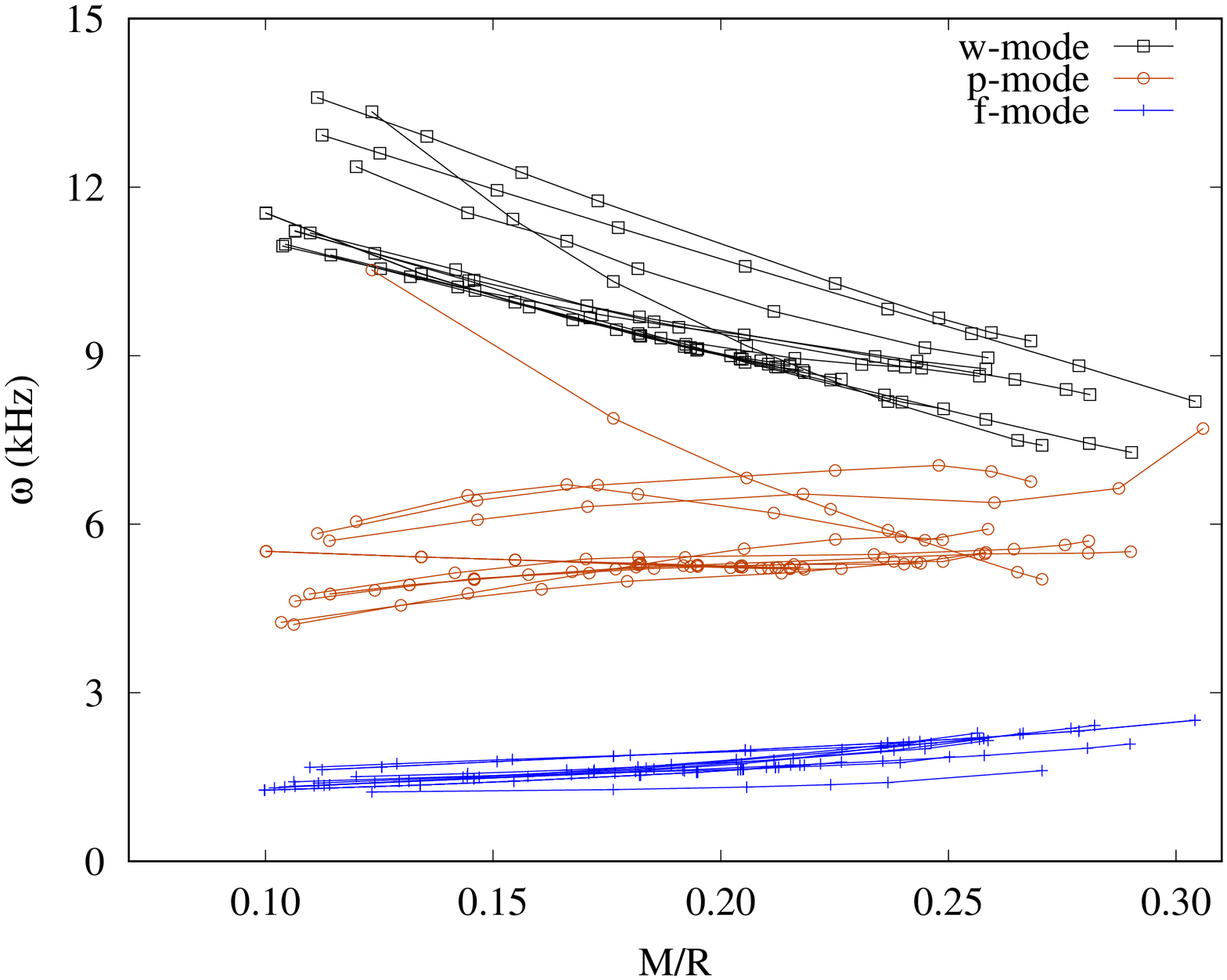}
\includegraphics[width=5.0cm, angle=270]{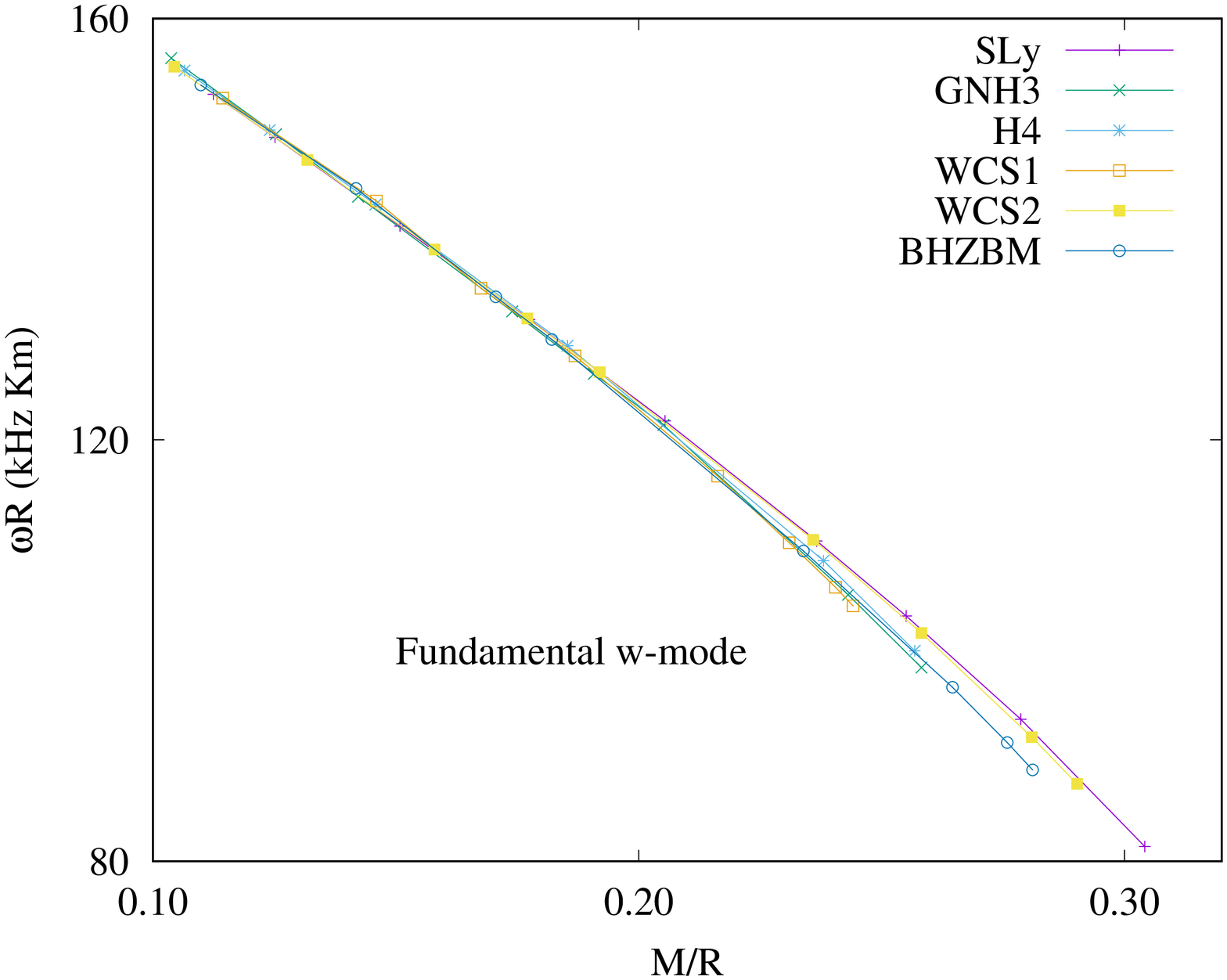}
}
\end{center}
\vspace{-0.5cm}
\caption{{$l=2$ polar QNMs: (a) frequency (in kHz) vs compactness $M/R$ for the fundamental f-mode, the pressure p-mode, and the space-time w-mode for several equations of state; (b) \textit{universal relation} for the scaled frequency $\omega R$ vs the compactness $M/R$ for the w-mode \cite{Blazquez-Salcedo:2013jka}.}} 
\label{fig2}
\end{figure}

As noticed early on \cite{Andersson:1996pn,Andersson:1997rn}, also the QNMs of neutron stars exhibit \textit{universal relations}. 
For better illustration, we show a set of $l=2$ polar QNMs in Fig.~\ref{fig2} \cite{Blazquez-Salcedo:2013jka}. 
These are the lowest mode, called the fundamental f-mode, the first pressure p-mode and the first space-time w-mode.
As seen in Fig.~\ref{fig2}(a), where we exhibit the frequency $\omega_R$ versus the compactness $M/R$, these modes typically exhibit a considerable dependence on the equation of state.
Fig.~\ref{fig2}(b) then demonstrates for the w-mode that by scaling the frequency with the radius a \textit{universal relation} results.
\textit{Universal relations} arise also for the axial modes.

\subsection{Neutron Stars beyond General Relativity}

Let us now consider the properties of neutron stars in some alternative gravities and begin with a set of
STTs, where in addition to the tensor a gravitational scalar degree of freedom is introduced \cite{Brans:1961sx,Damour:1992we,Fujii:2003pa}.
When the STT action is formulated in the physical Jordan frame, the nuclear matter is minimally coupled to the Jordan frame metric $\tilde g_{\mu\nu}$.
Subsequent transformation to the Einstein frame then leads to a coupling of the scalar to the nuclear matter since
the Jordan frame metric and the Einstein frame metric $g_{\mu\nu}$ are related by 
\begin{equation}
    \tilde g_{\mu\nu} = A^2(\varphi) g_{\mu\nu} \ ,
\end{equation}
where $\varphi$ denotes the scalar field in the Einstein frame, and the coupling function $A(\varphi)$ depends on the respective STT.

Whereas the old Brans-Dicke theory is basically ruled out by observations in the solar system by now, other STTs are still interesting candidate theories, that may be constrained by observational data from compact objects.
Therefore theoretical studies of the consequences of such theories for neutron star properties and gravitational wave emission are of importance, in particular, since the existence of \textit{universal relations} may help largely to pinpoint effects of these alternative gravities, when the \textit{universal relations} vary distinctly with the coupling constants of these theories.

We first address STTs that allow for matter induced spontaneous scalarization of neutron stars \cite{Damour:1993hw}.
To understand this phenomenon, we inspect the coupled set of field equations in the Einstein frame 
\begin{eqnarray}
R_{\mu\nu} &=& 2 \partial_\mu \varphi \partial_\nu \varphi 
+ 8 \pi G \left( T_{\mu\nu} - \frac{1}{2} T g_{\mu\nu} \right) \ , \label{einmat} \\
\nabla_\mu \nabla^\mu \varphi + 4\pi G \alpha(\varphi) T &=&
\nabla_\mu \nabla^\mu \varphi - m^2_{\rm eff} \varphi = 0 \ , \label{phi_mat}
\end{eqnarray}
with stress energy tensor $T_{\mu\nu}$, trace $T$, and $\alpha=d(\ln A(\varphi))/d\varphi$, and the scalar equation has been rewritten in terms of an effective mass $m_{\rm eff}^2$.

For spontaneous scalarization to occur, the coupling function $A(\varphi)$ has to be chosen such that for $\varphi=0$ also $\alpha=0$.
In that case the GR equations are satisfied, when the scalar field vanishes, and the GR neutron star solutions remain solutions of the STT.
However, in addition to the GR solutions also scalarized neutron star solutions can arise, when the scalar field equation develops a tachyonic instability.
For instance, the choice \cite{Damour:1993hw}
\begin{equation}
    A=\exp{(\frac{1}{2} \beta_0 \varphi^2)} \ , \ \ \alpha = \beta_0 \varphi
    \label{alpha}
\end{equation}  
allows for a tachyonic instability, $m_{\rm eff}^2 = - 4\pi G \beta_0 T < 0$, when both the coupling constant $\beta_0$ and the trace $T=3p-\rho$ are negative, and form a sufficiently strong source for the scalar field. 

\begin{figure}[ht]
\begin{center}
\mbox{
\includegraphics[width=6.6cm, angle=0]{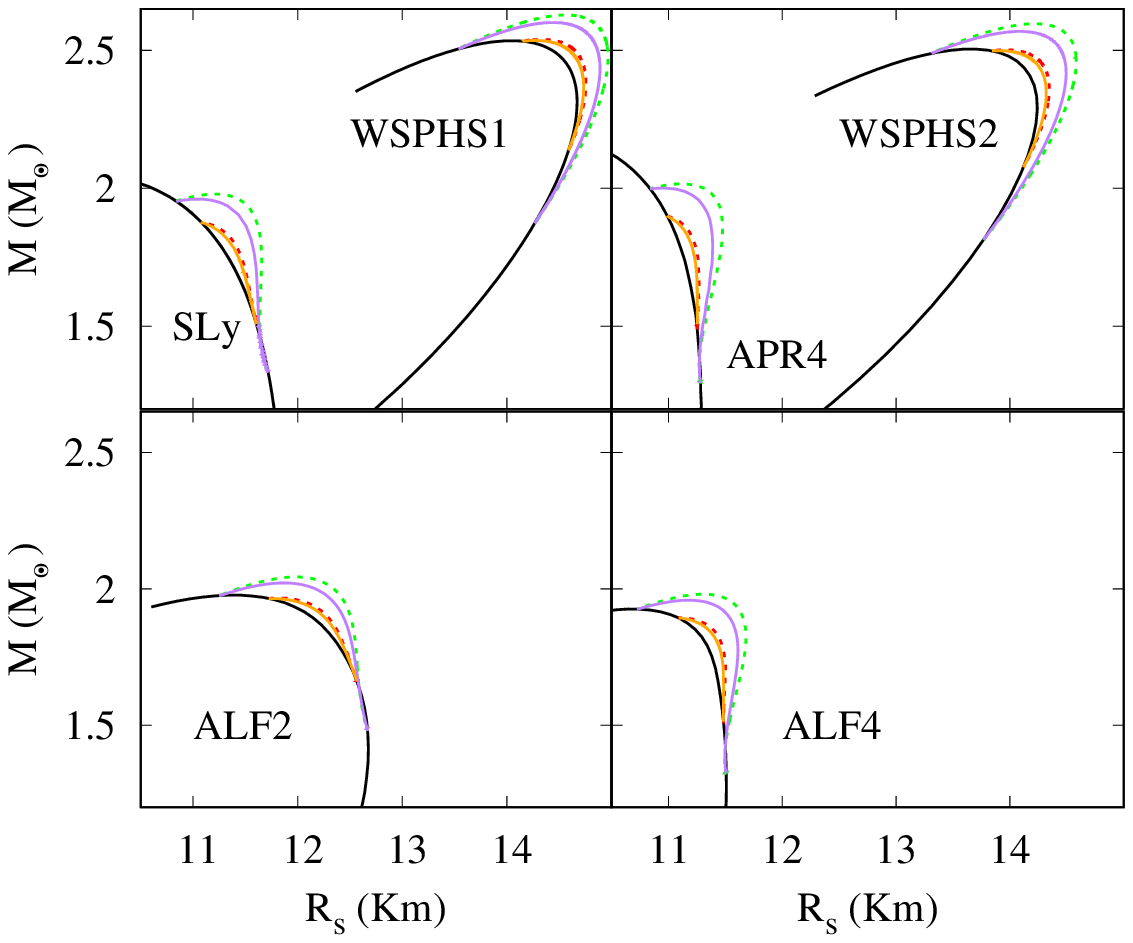}
\includegraphics[width=6.6cm, angle=0]{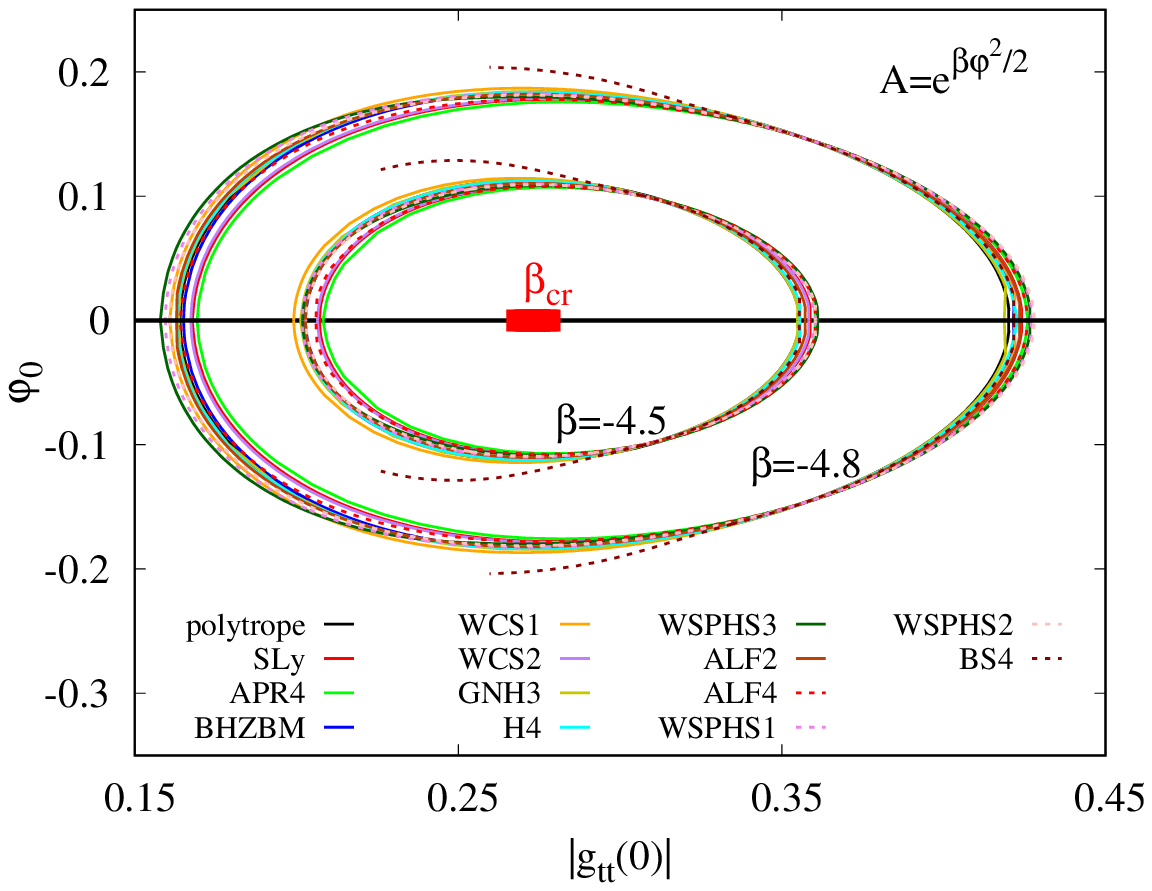}
}
\end{center}
\vspace{-0.5cm}
\caption{{Spontaneous scalarization of neutron stars: (a) mass--radius relation for several equations of state for two values of $\beta_0$, also shown is the GR relation;
(b) \textit{universal relation} for the value of the scalar field at the center $\varphi(0)$ vs the $tt$ component of the metric at the center $g_{tt}(0)$. {$\beta_{\rm cr}$ denotes the onset of scalarization}
 \cite{AltahaMotahar:2017ijw}.}}
\label{fig3}
\end{figure}

Typically, neutron stars possess $T<0$, therefore the coupling constant $\beta_0$ should be sufficiently negative to achieve spontaneous scalarization.
We note, however, for later reference, that the pressure-dominated core of neutron stars with very high densities may lead to regions with $T>0$, and therefore spontaneous scalarization is also possible for such stars if the coupling constant is sufficiently positive, $\beta_0>0$ \cite{Mendes:2014ufa,Mendes:2016fby}.
Returning to $\beta_0<0$, we exhibit in Fig.~\ref{fig3} the effect of matter-induced scalarization for several equations of state \cite{AltahaMotahar:2017ijw}.
Fig.~\ref{fig3}(a) shows the mass--radius relation for several equations of state for two values of $\beta_0$ together with the GR relation.
Moving along a family of stars the GR neutron stars develop an unstable mode when scalarization sets. 
In contrast, the scalarized neutron stars are stable and represent the preferred configurations.
Scalarization is the stronger the more negative the coupling constant $\beta_0$ becomes.
Fig.~\ref{fig3}(b) illustrates a kind of \textit{universal relation} that arises when the value of the scalar field at the center $\varphi(0)$ is considered versus the gravitational potential at the center (as contained in $g_{tt}(0)$).

In such STTs the presence of the scalar degree of freedom would lead to an appreciable amount of dipolar radiation and thus to a more rapid orbital decay, while in GR only quadrupole (and higher multipole) radiation can be emitted.
{Therefore observational data from binary pulsars have been used to put constraints on the parameters of these theories  
(see e.g.~\cite{Alsing:2011er,Freire:2012mg}), 
leading recently even to the exclusion of the possibility of spontaneous scalarization for the above discussed class of massless STTs \cite{Zhao:2022vig}.
This conclusion can be avoided, however,} when the scalar field has a larger mass, since the dipolar radiation is effectively shut off when the Compton wavelength of the scalar field is small in comparison to the orbital separation of a binary system \cite{Ramazanoglu:2016kul}.
The effects of scalar field masses and of scalar self-interaction have, for instance, been studied in \cite{Yazadjiev:2016pcb,Doneva:2016xmf,AltahaMotahar:2019ekm},
where also \textit{universal relations} were obtained, that were distinct from those of GR.

We next consider neutron stars in a particular $f(R)$ theory,
\begin{equation}
    f(R)=R+aR^2 \ .
    \label{R2}
\end{equation}
Mathematically $f(R)$ theories are equivalent to STTs, and can be transformed to the Einstein frame.
The theory (\ref{R2}) then involves a coupling function $A(\varphi)$ and a potential $V(\varphi)$, 
\begin{equation}
    A(\varphi)= e^{-\frac{1}{\sqrt{3}}\varphi}  \ , \ \ \ V(\varphi)=\frac{3m_{\varphi}^2}{2} \big(1- e^{-\frac{2\varphi}{\sqrt{3}}}\big)^2 \ ,
\end{equation}
where the scalar field mass $m_\varphi$ is related to the coupling constant $a$ by $m_\varphi=1/\sqrt{6a}$ \cite{Yazadjiev:2014cza,Staykov:2014mwa}.
Then a coupling constant of $a=1$ corresponds to a mass of $m_\varphi=1.08$ neV, while $a=10000$ to $m_\varphi=0.0108$ neV.
Such a mass range represents an interesting observational window \cite{Ramazanoglu:2016kul} inviting theoretical studies \cite{Yazadjiev:2014cza,Staykov:2014mwa,Yazadjiev:2015zia,Doneva:2015hsa,Blazquez-Salcedo:2018qyy,Blazquez-Salcedo:2020ibb,Blazquez-Salcedo:2021exm}.

\begin{figure}[ht]
\begin{center}
\mbox{
\includegraphics[width=6.9cm, angle=270]{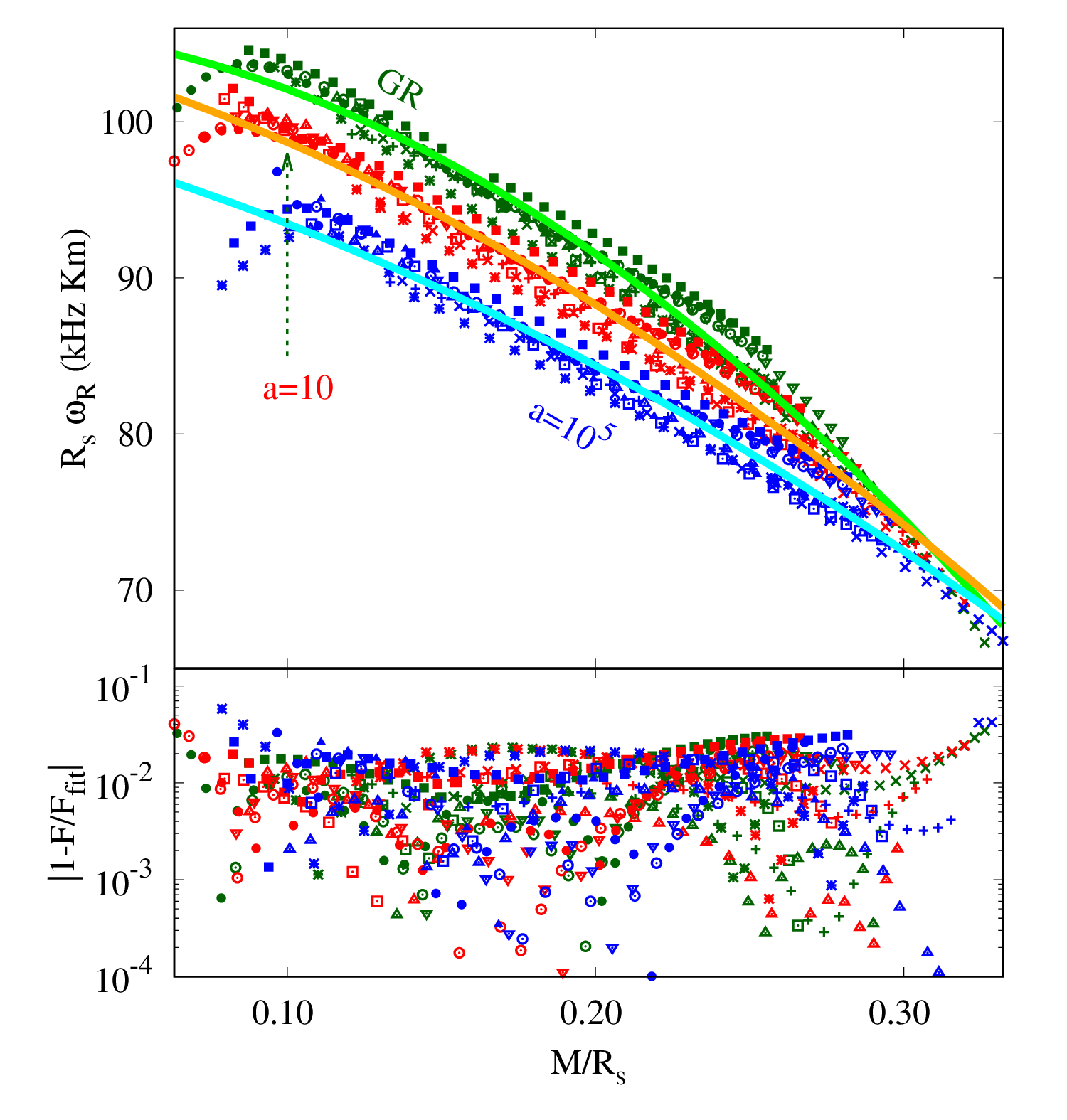}
\includegraphics[width=6.9cm, angle=270]{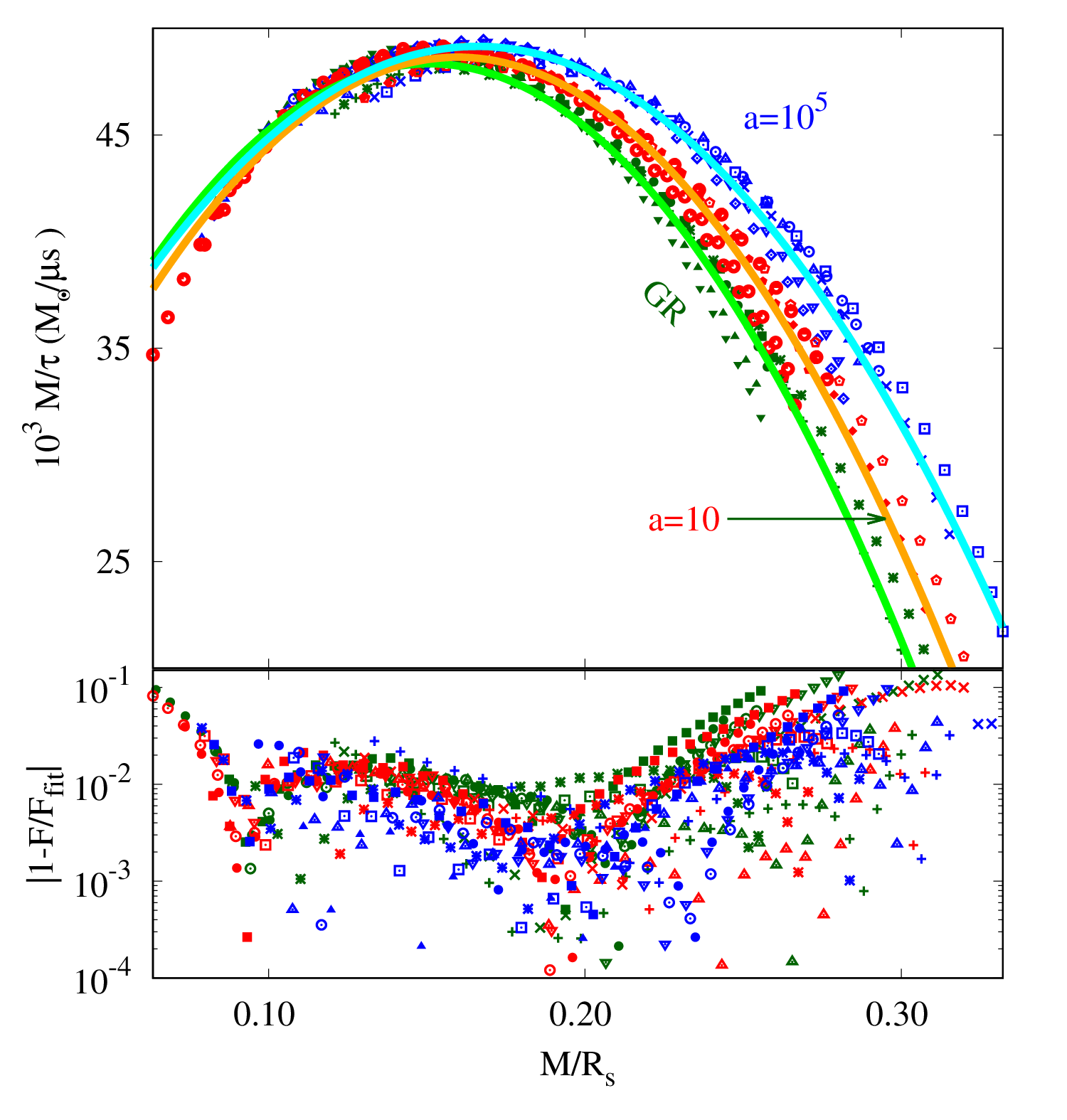}
}
\end{center}
\vspace{-0.5cm}
\caption{{\textit{Universal relations} for axial $l=2$ QNMs of neutron stars in $R^2$ gravity: (a) scaled frequency $\omega_R R$ vs compactness $M/R$ for $a=10$ and $a=10^5$; (b) scaled damping time $M/\tau$ vs $M/R$. Also shown are the deviations from the respective best fits, as well as the GR modes for comparison \cite{Blazquez-Salcedo:2018qyy}.}}
\label{fig4}
\end{figure}

Concerning \textit{universal relations}, the $I$-$Q$ relations have been obtained for $a=10^4$ not only for the slowly rotating case but also for rapidly rotating neutron stars, where the additional dependence on the dimensionless angular momentum $J/M^2$ leads to a \textit{universal relation} surface \cite{Doneva:2015hsa} as in GR \cite{Pappas:2013naa,Chakrabarti:2013tca}.
However, for such a value of the scalar mass, the GR relation and the $R^2$ relation differ distinctly \cite{Doneva:2015hsa}.
Of interest are also the \textit{universal relation} for QNMs.
We exhibit examples for axial $l=2$ modes in Fig.~\ref{fig4}, where the scaled frequency $\omega_R R$ and the scaled damping time $M/\tau$ are exhibited versus the compactness $M/R$ for $a=10$ and $a=10^5$ and compared with the corresponding GR modes \cite{Blazquez-Salcedo:2018qyy}.

Axial modes do not involve perturbations of the matter and of the scalar field \cite{Blazquez-Salcedo:2018qyy}.
This is in contrast to polar modes, which exhibit a much richer spectrum \cite{Blazquez-Salcedo:2020ibb,Blazquez-Salcedo:2021exm}.
Moreover, the additional scalar degree of freedom allows for $\varphi$ modes, and the $l=0$ normal modes of GR become propagating QNMs in $R^2$ gravity. 
The detection of gravitational monopole or dipole radiation (i.e., $l<2$) would be a clear sign of the new gravitational degree of freedom.

\begin{figure}[ht]
\begin{center}
\mbox{
\includegraphics[width=5.0cm, angle=270]{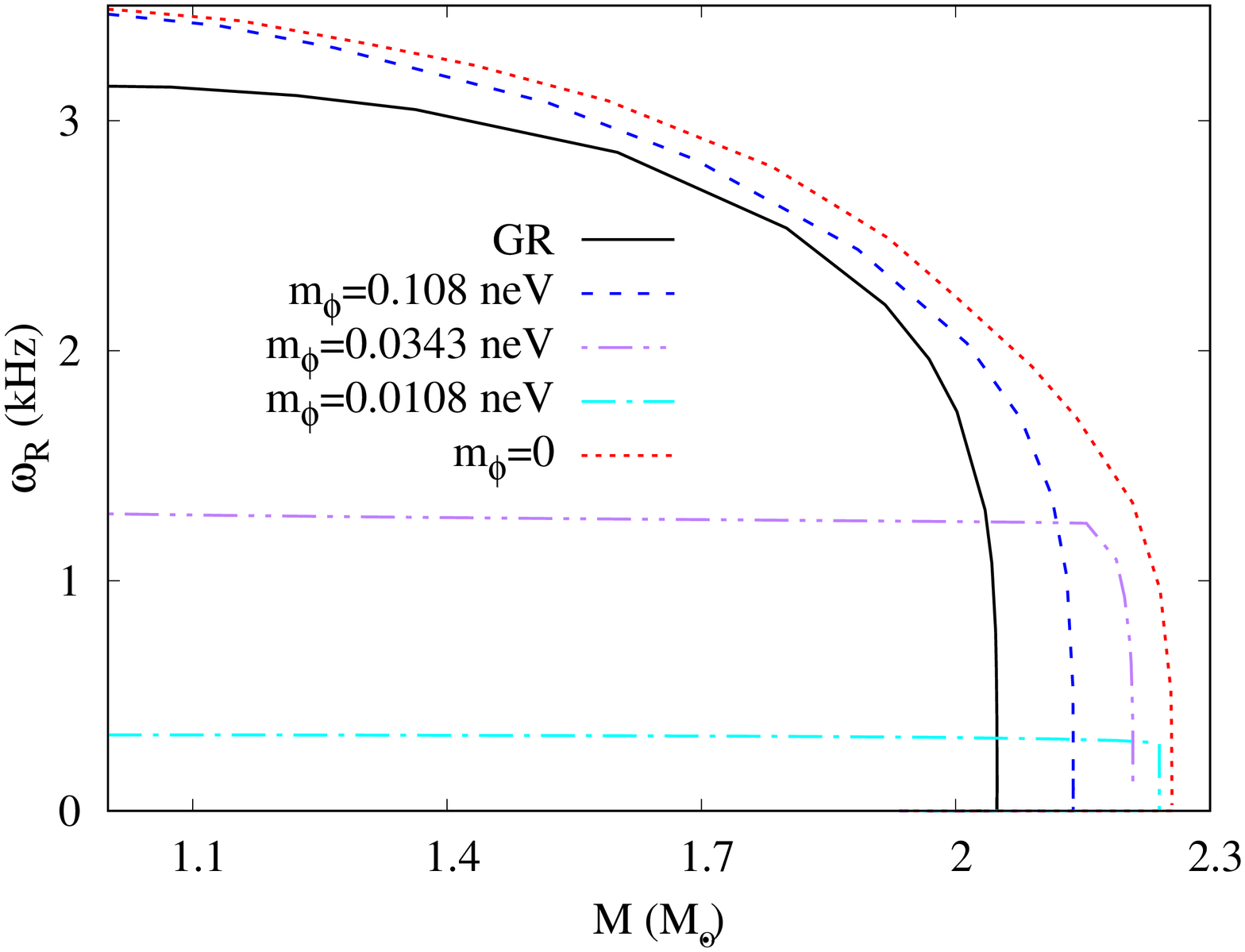}
\hspace{-.3cm}
\includegraphics[width=5.0cm, angle=270]{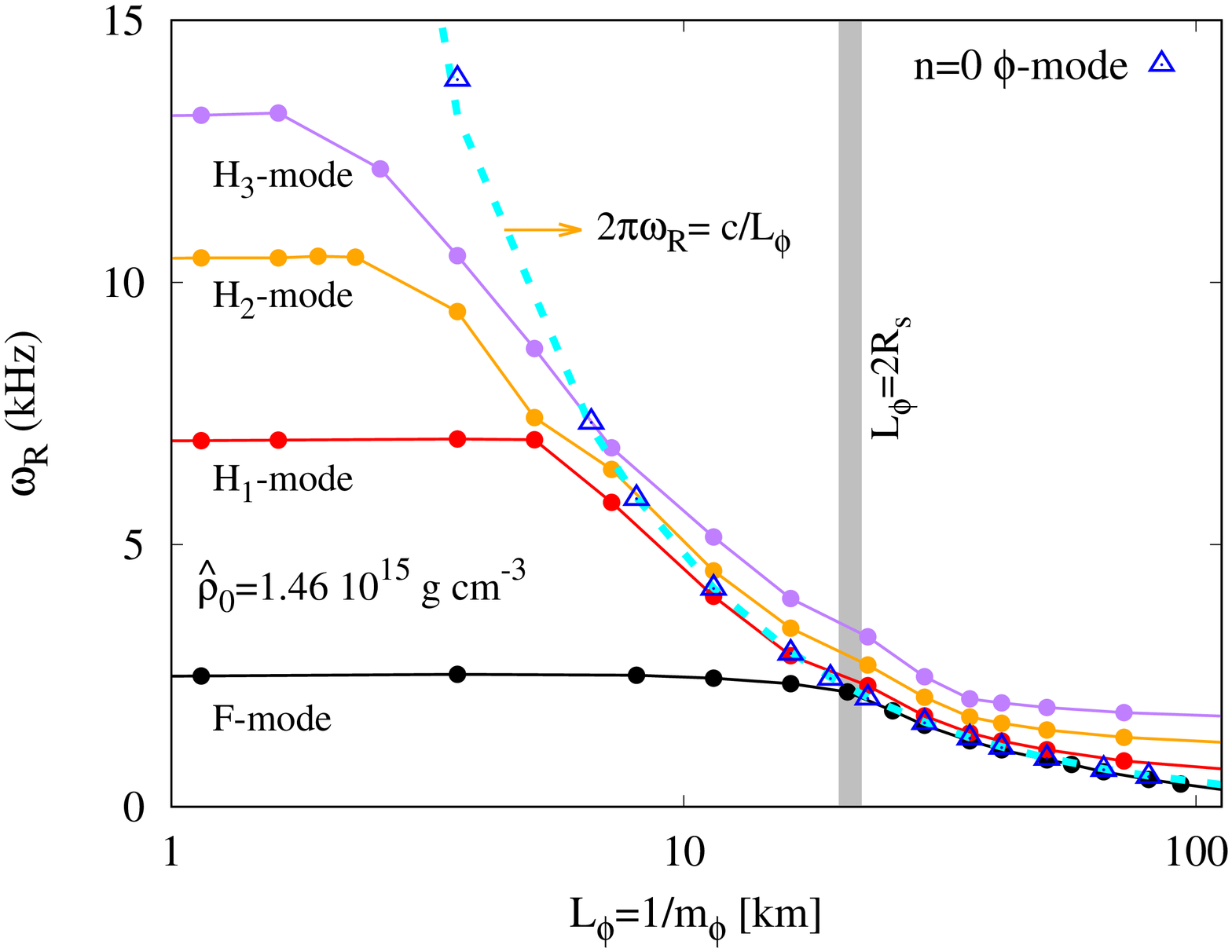}
}
\end{center}
\vspace{-0.5cm}
\caption{{Polar QNMs of neutron stars in $R^2$ gravity: (a) frequency $\omega_R$ vs mass $M$ of the $l=0$ F mode for several values of $a$, also shown are the GR values and the massless case; (b) frequency $\omega_R$ vs scalar Compton wavelength $L_\varphi$ for the F mode, H modes and $\varphi$ mode for a given neutron star central density. The equation of state is fixed \cite{Blazquez-Salcedo:2020ibb}.}}
\label{fig5}
\end{figure}

Fig.~\ref{fig5}(a) exhibits for a prescribed equation of state the dependence of the frequency of the $l=0$ F mode on the mass $M$ of the neutron stars for several values of the scalar mass $m_\varphi$. It also compares with the massless case and with GR.
In GR the F mode is a normal mode, that becomes a zero mode at the neutron star maximum mass, turning into an unstable mode beyond the maximum mass.
In contrast, in $R^2$ gravity the additional degree of freedom leads for the stable neutron stars to QNMs, that possess a very small imaginary part of the eigenvalue $\omega$, making them ultra long-lived \cite{Blazquez-Salcedo:2020ibb}.

The F mode is shown in Fig.~\ref{fig5}(b) as a function of the scalar Compton wavelength for a given neutron star central density.
Also shown are the excited H modes and the scalar $\varphi$ mode.
Interestingly, the scale of the frequency is set by (multiples of) the size of the star as long as the Compton wavelength is small, whereas for larger Compton wavelength the scale is set by the scalar field mass, as indicated by the curve $2 \pi \omega_R = m_\varphi$ \cite{Blazquez-Salcedo:2020ibb}.
\textit{Universal relations} for the polar modes can be obtained as well, as preliminary results including the $l=1$ and $l=2$ modes suggest.

As a final example of alternative gravities we here briefly address a STT with Chameleon screening \cite{Brax:2012gr}.
We therefore consider neutron stars in a STT with Einstein frame coupling function $A(\varphi)$ and potential $V(\varphi)$  
\begin{equation}
    A(\varphi)= e^{\alpha_0 \varphi} \ , \ \ \
V(\varphi)=\frac{ \Lambda^{n+4}}{\varphi^n} \ , \ \ \ n=1 \ ,
\label{cham}
\end{equation}
where $\alpha_0$ is the coupling constant, and $\Lambda$ sets the chameleon energy scale \cite{deAguiar:2020urb,Dima:2021pwx}.
In Chameleon screening the scalar field obtains an effective mass whose value is given by the surrounding matter density
\begin{equation}
    m^2_{\rm eff} = \frac{2 \Lambda^{5}}{\varphi^3} - 4 \alpha_0^2 e^{ 4 \alpha_0 \varphi} \tilde T \ ,
\end{equation}
where $\tilde T$ denotes the trace of the stress energy tensor of the matter in the Jordan frame.

Since the scalar field satisfies a Klein-Gordan equation for an effective potential
\begin{equation}
    V(\varphi)_{\rm eff} = \frac{ \Lambda^{5}}{\varphi} -\frac{1}{4} e^{ 4 \alpha_0 \varphi} \tilde T \ ,
\end{equation}
the scalar field will assume the minimum value of the effective potential, thus depending strongly on the surrounding matter.
The Chameleon field is then short-ranged in regions of high density and long-ranged on cosmological scales.
For such a theory to remain viable, however, also compact objects like neutron stars should be well described and in accordance with observations.

As shown in \cite{deAguiar:2020urb,Dima:2021pwx} such a Chameleon theory leads to stable neutron stars.
This remains the case even when the trace of the stress energy tensor changes sign and the neutron stars develop a pressure dominated core with $T>0$, and therefore a partial unscreening of the scalar field arises in the interior \cite{deAguiar:2020urb,Dima:2021pwx}.
Predictions concerning the observability of scalar radiation from neutron stars in such a Chameleon theory must rely, however, on extrapolation, since reliable calculations could be performed only for $\Lambda=175$ GeV, $\tilde \rho_\infty = 6.5 \times 10^{+10}$ g/cm$^3$ (where $\tilde \rho_\infty$ is the density outside the star), with the physical values of these quantities differing by many orders of magnitude, $\Lambda=\,  2.4$ meV, $\tilde \rho_\infty = 1.67 \times 10^{-20}$ g/cm$^3$. 
Extrapolation then predicts a peak in the mHz band in the sensitivity range of LISA, but far below its sensitivity \cite{Dima:2021pwx}.

\section{Black Holes}

\subsection{Black Holes in General Relativity}

The Kerr black hole of GR is considered of utmost astrophysical relevance \cite{Kerr:1963ud,Teukolsky:2014vca}.
It represents a fascinating family of asymptotically flat vacuum solutions that are regular on and outside their event horizon, and that are uniquely specified by their mass $M$ and angular momentum $J$  (see e.g. \cite{Chrusciel:2012jk}).
Kerr black holes carry \textit{no hair}.
Their higher mass moments $M_l$ and current moments $S_l$ are simply given in terms of the mass $M_0=M$ and the angular momentum $S_1=J$ \cite{Geroch:1970cd,Hansen:1974zz}
   \begin{equation}                                              M_l + i S_l = M \left( i \frac{J}{M} \right)^l \ , \end{equation} 
in which odd mass moments and even current moments vanish. 
The \textit{Kerr paradigm} supposes, that indeed all astrophysical black holes are well described by Kerr black holes \cite{Bambi:2011zs}.
Testing the \textit{no-hair} hypothesis and the \textit{Kerr paradigm} with current and future observations is clearly a central issue \cite{Cardoso:2016ryw,Cardoso:2019rvt}.

EHT observations of the shadow of the supermassive black hole at the center of the elliptic galaxy M87 are in good agreement with theoretical predictions obtained with full-fledged general relativistic magneto-hydrodynamics (GRMHD) calculations \cite{EventHorizonTelescope:2019dse}.
Of course, the shadow of a Kerr black hole itself and the dependence of its shape on the angular momentum has been known since long \cite{Bardeen:1973tla}.

On the other hand, gravitational wave observations from merger events seen by LIGO/VIRGO employ standard GR templates based on Kerr black holes in their analyses \cite{LIGOScientific:2016aoc,LIGOScientific:2017vwq}.
Whereas the merger itself requires numerical relativity, the ringdown involves the QNMs of Kerr black holes, that are well-known (see e.g. \cite{Berti:2009kk,Konoplya:2011qq}).
An interesting properties of these Kerr QNMs is their \textit{isospectrality}, i.e., the eigenvalues of polar and axial perturbations are identical.

Besides the Kerr black holes and their static limit, the Schwarzschild black holes (as well as the charged Kerr-Newman and Reissner-Nordstr\"om black holes), GR features also asymptotically flat black holes with hair (see, e.g., \cite{Volkov:1998cc,Kleihaus:2016rgf,Herdeiro:2014goa}).
While the hairy black holes with \textit{Standard model} fields \cite{Volkov:1998cc,Kleihaus:2016rgf} appear to be astrophysically irrelevant, this could be different for Kerr black holes with scalar hair \cite{Herdeiro:2014goa} or Proca hair \cite{Herdeiro:2016tmi}.
While the scalar and Proca fields needed would be new types of particles beyond the \textit{Standard Model}, the corresponding rotating black holes represent interesting alternatives to the Kerr black holes and so far viable astrophysical candidates.

\subsection{Black Holes beyond General Relativity}

The STTs dicussed above, that possess matter-induced spontaneously scalarized neutron stars, do not support scalarized black holes.
The \textit{no-hair} theorems of GR forbid such solutions, and only Schwarzschild and Kerr black holes prevail.
These remain also solutions of $f(R)$ theories, as evident in the Einstein frame.
The presence of the additional degree of freedom will, however, modify their QNM spectra and thus possibly implicate observable effects \cite{Berti:2005ys}.

Here we would like to address another type of alternative gravity that includes a scalar degree of freedom and leads to black holes with scalar hair.
In particular, we focus on a set of higher curvature theories, Einstein-scalar-Gauss-Bonnet (EsGB) theories \cite{Zwiebach:1985uq,Gross:1986mw,Metsaev:1987zx}, where the Einstein action is supplemented by the Gauss-Bonnet (GB) invariant
\begin{equation}
    R^2_{\rm GB} = R_{\mu\nu\rho\sigma} R^{\mu\nu\rho\sigma}
- 4 R_{\mu\nu} R^{\mu\nu} + R^2 \ ,
\end{equation}
coupled to a scalar field $\varphi$ with a coupling function $f(\varphi)$, and also a scalar kinetic term and, possibly, a scalar potential.
The action leads to second order field equations without ghosts and therefore represents a certain type of Horndeski theory \cite{Horndeski:1974wa}.

Depending on the coupling function $f(\varphi)$, different types of scalarization arise, as seen from the field equations.
Let us therefore take a closer look at the scalar field equation 
\begin{equation}
    \nabla_\mu \nabla^\mu \varphi + \frac{df}{d\varphi} R^2_{\rm GB}  =0 \ .
    \label{scalar}
\end{equation}
GR black hole solutions can not remain solutions, when $\frac{df}{d\varphi} \ne 0$, since the scalar field would have a non-vanishing source term, and the resulting finite scalar field would contribute in the Einstein equations.
In this case only hairy black holes arise.
When, however, $\frac{df}{d\varphi} = 0 $ for $\varphi=0$, then the GR black holes remain solutions of the theory, but additional scalarized solutions may arise. 

\begin{figure}[ht]
\begin{center}
\mbox{
\includegraphics[width=.5\textwidth, angle =0]{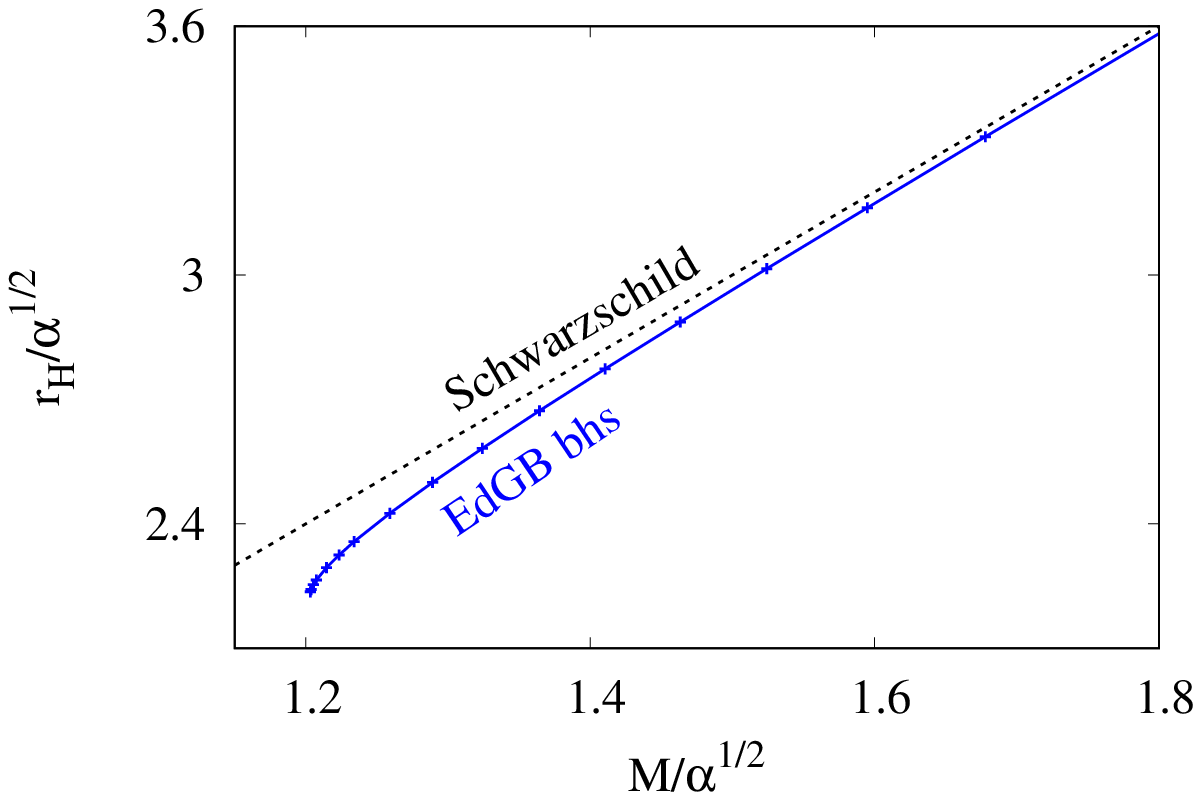}
\includegraphics[width=.5\textwidth, angle =0]{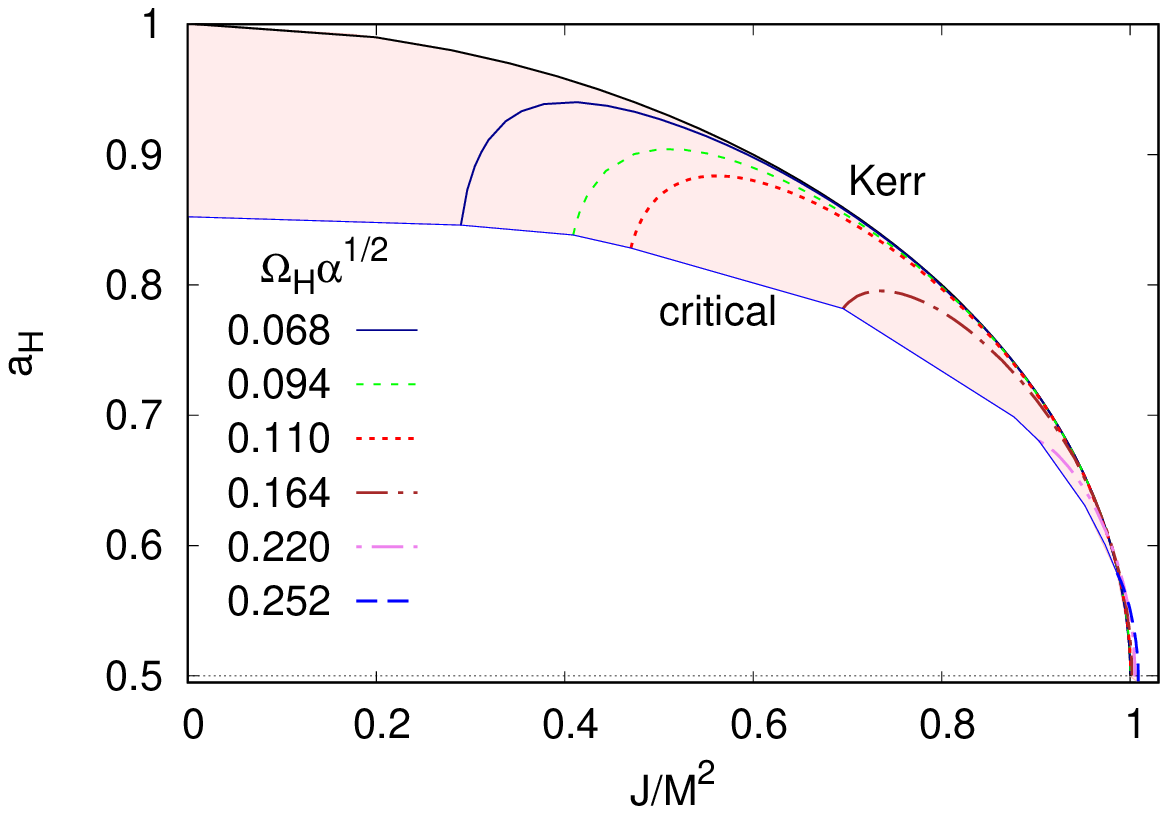}
}
\end{center}
\vspace{-0.5cm}
\caption{{Dilatonic GB black holes: (a) scaled event horizon radius $r_H/\sqrt{\alpha}$ vs scaled mass $M/\sqrt{\alpha}$ of static black holes, compared to Schwarzschild; (b) scaled horizon area $a_H=A_H/16 \pi M^2 $ vs scaled angular momentum $j=J/M^2$ of rotating black holes, delimited by static and Kerr black holes, and critical solutions. {Also shown are curves of fixed scaled horizon angular velocity $\Omega_H \alpha^{1/2}$} \cite{Kleihaus:2011tg}.}}
\label{fig6}
\end{figure}

When the scalar field is considered to be a dilaton, as motivated by string theory, the coupling function $f(\varphi)=\alpha \exp{(-\varphi)}/4$ is of the first type, and only hairy black holes arise \cite{Kanti:1995vq,Torii:1996yi,Guo:2008hf,Kleihaus:2011tg,Kleihaus:2014lba,Kleihaus:2015aje}.
Fig.~\ref{fig6}(a) shows the scaled event horizon radius $r_H/\sqrt{\alpha}$ versus the scaled mass $M/\sqrt{\alpha}$ of the family of static dilatonic black hole solutions, comparing them with the Schwarzschild solutions.
Clearly, at fixed coupling constant $\alpha$, large dilatonic black holes approach the Schwarzschild black holes.
Smaller dilatonic black holes deviate significantly, however, and cease to exist beyond a minimal mass.
The reason is of purely theoretical nature.
The horizon expansion of the dilaton involves a square root, $\sqrt{1-6\alpha^2 e^{2 \phi_H}/r_H^4}$, whose radicand vanishes when the minimal mass is reached \cite{Kanti:1995vq}.

When these black holes are set into rotation, the dilatonic generalizations of the Kerr black holes are obtained \cite{Kleihaus:2011tg}.
Their domain of existence is exhibited in Fig.~\ref{fig6}(b), where the scaled horizon area $a_H=A_H/16 \pi M^2 $ is shown versus the scaled angular momentum $j=J/M^2$.
The left boundary corresponds to the static dilatonic black holes, the upper boundary to the Kerr black holes, and the lower boundary to the critical solutions (where a radicand vanishes).
Interestingly, these dilatonic black holes can slightly violate the Kerr bound, $J/M^2=1$. 
As expected from the static case, the area of the rotating dilatonic black  holes is smaller than the area of Kerr black holes with the same angular momentum.
The shadow is only slightly smaller than the shadow of comparable Kerr black holes \cite{Cunha:2016wzk}.
The entropy of dilatonic black  holes is, however, larger since the entropy receives an extra contribution from the GB term \cite{Wald:1993nt}.

\begin{figure}[ht]
\begin{center}
\mbox{
\includegraphics[width=4.5cm, angle =270]{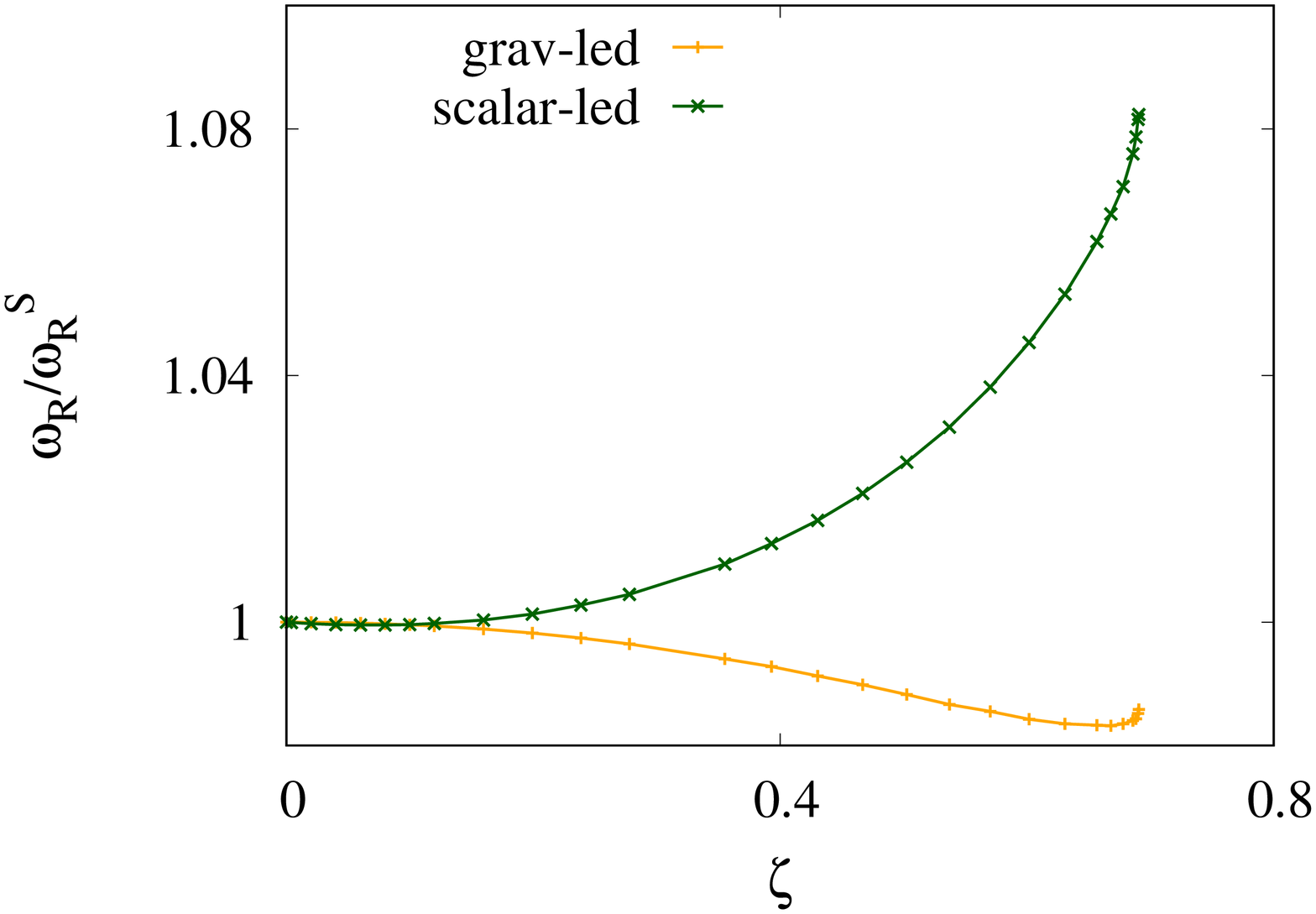}
\includegraphics[width=4.5cm, angle =270]{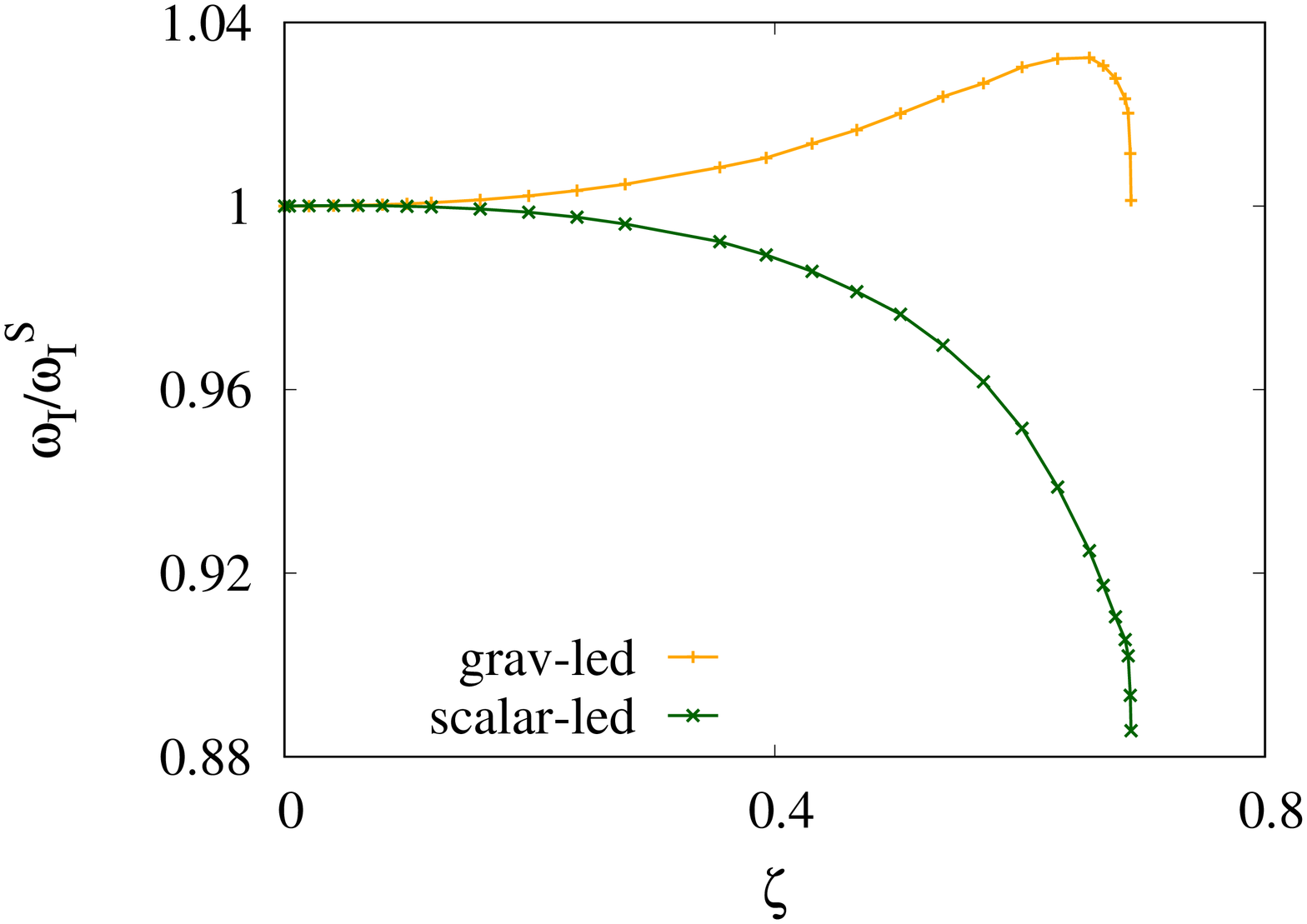}
}
\end{center}
\vspace{-0.5cm}
\caption{{Polar $l=2$ QNMs of dilaton GB black holes: (a) scaled frequency $\omega_R/\omega_R^S$ vs dimensionless coupling constant $\zeta=\alpha/M^2$; (b) scaled decay rate $\omega_I/\omega_I^S$ vs $\zeta=\alpha/M^2$.
Shown are the scalar-led and grav-led modes. %
The scaling is w.r.t.~Schwarzschild \cite{Blazquez-Salcedo:2016enn}.}}
\label{fig7}
\end{figure}

QNMs of static dilatonic GB black holes have been considered widely
\cite{Kanti:1997br,Pani:2009wy,Ayzenberg:2013wua,Blazquez-Salcedo:2016enn,Blazquez-Salcedo:2017txk,Konoplya:2019hml,Zinhailo:2019rwd}.
The scalar field allows for $l=0$ and $l=1$ radiation, and also for scalar-led higher $l$ modes in addition to the grav-led modes, that reduce to the GR modes for vanishing GB coupling.
As an example we exhibit in Fig.~\ref{fig7} the QNM eigenvalue for the polar $l=2$ scalar-led and grav-led modes versus the strength of the GB coupling constant. The latter is scaled by the mass, $\zeta=\alpha/M^2$. The frequency $\omega_R$ and the decay rate $\omega_I$ are scaled by the respective Schwarzschild values \cite{Blazquez-Salcedo:2016enn}.
Close to the critical solution the deviations from Schwarzschild become more pronounced.
The presence of the scalar field breaks isospectrality of the modes.
Recent developments concern the perturbative inclusion of rotation for QNMs \cite{Pierini:2021jxd}, and time evolution of the gravitational wave signal \cite{Shiralilou:2020gah,Shiralilou:2021mfl}.

\begin{figure}[ht]
\begin{center}
\mbox{
\includegraphics[width=5.0cm, angle =270]{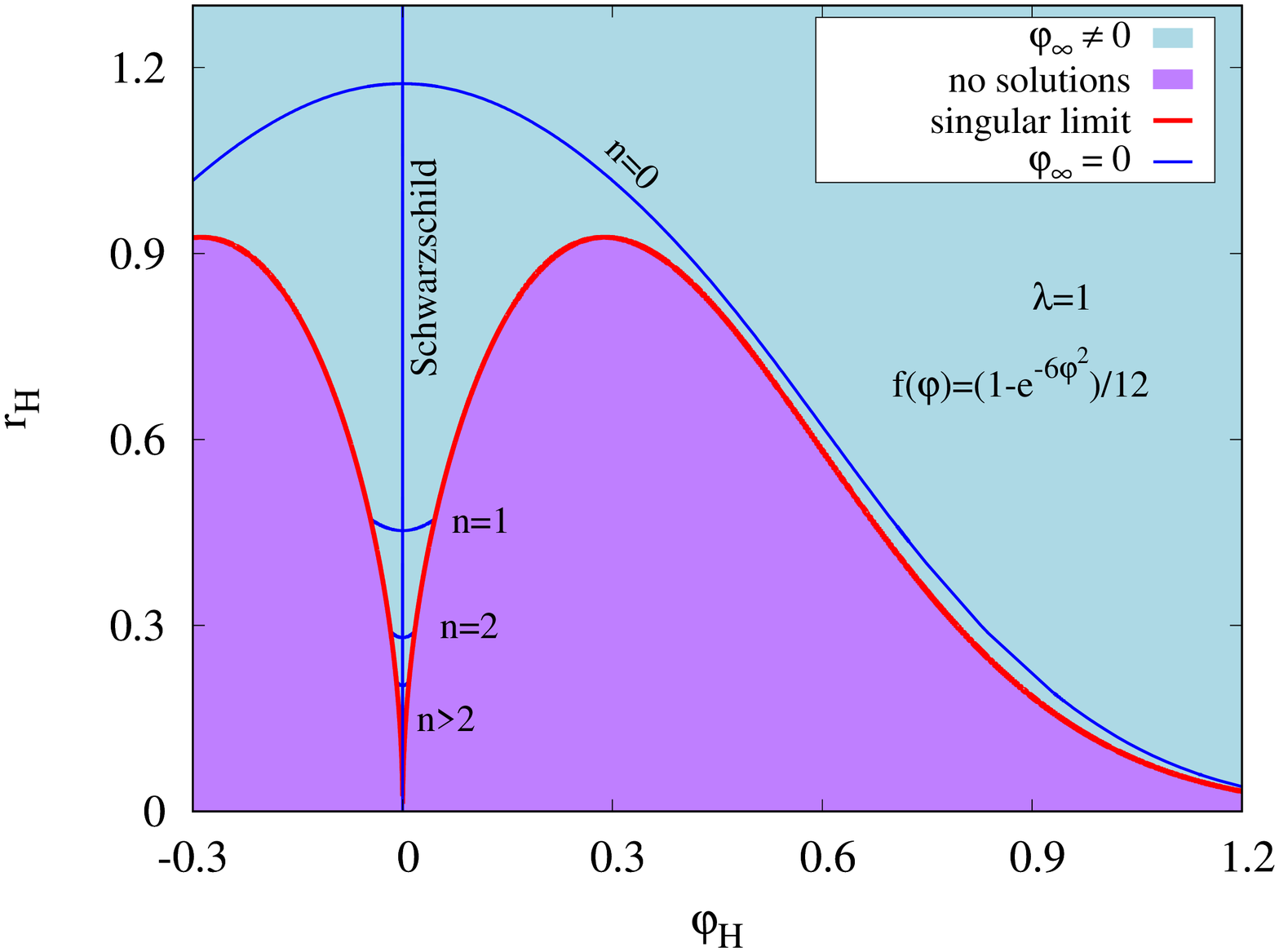}
\includegraphics[width=5.0cm, angle =270]{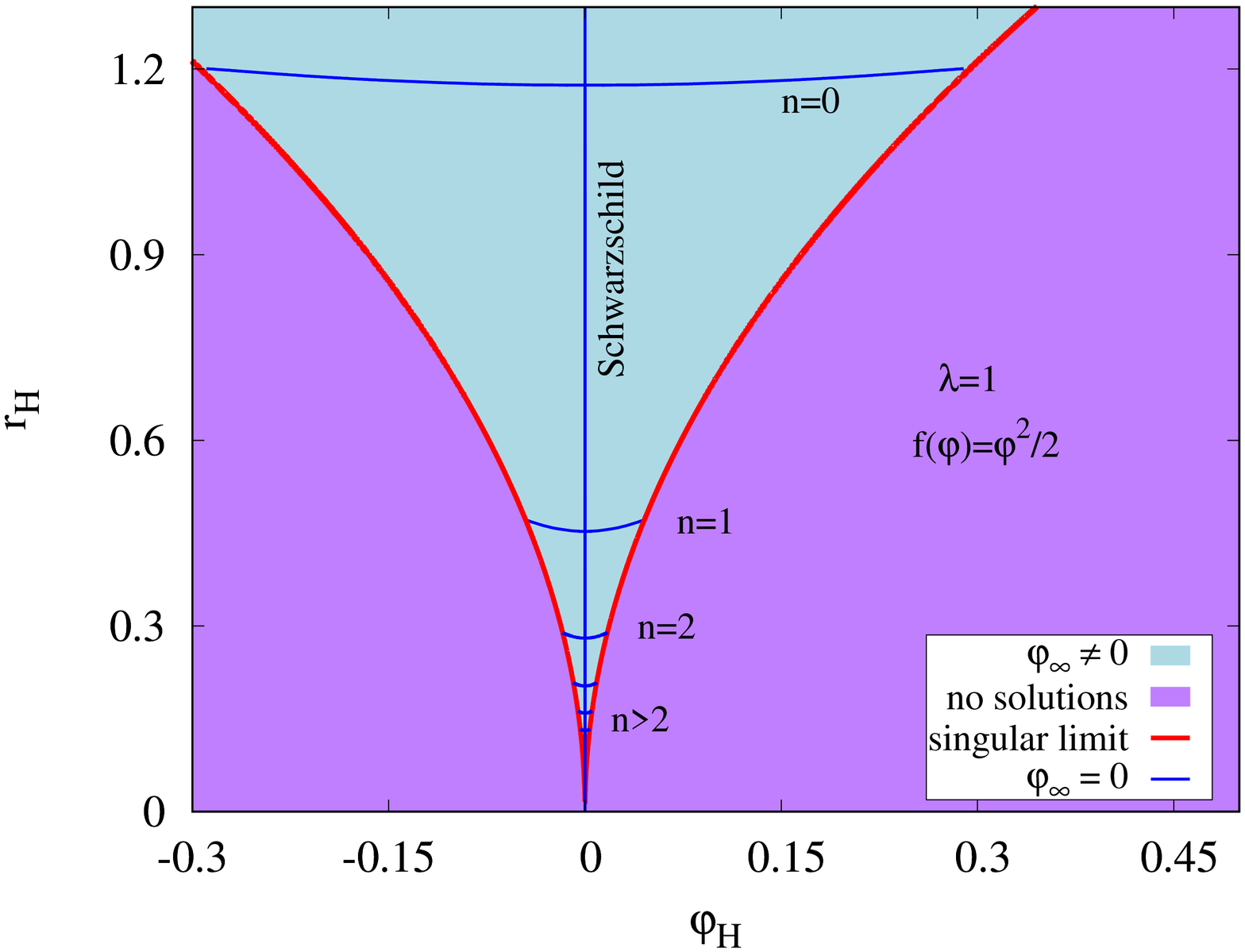}
}
\end{center}
\vspace{-0.5cm}
\caption{{Domain of existence of static spontaneously scalarized EsGB black holes: (a) horizon radius $r_H$ vs horizon scalar field $\varphi_H$ at fixed GB coupling $\lambda=1$ for exponential coupling function; (b) same for quadratic coupling function.
Shown are the fundamental mode $n=0$, and the lowest radially excited modes $n=1$ and 2. The colors indicate forbidden regions
\cite{Blazquez-Salcedo:2020ibb}.}}
\label{fig8}
\end{figure}

We now turn to the second type of coupling function $f(\varphi)$, and require that the coupling function vanishes for vanishing scalar field.
The resulting EsGB theory then retains the Schwarzschild and Kerr black holes as solutions.
When the coupling is such, that the scalar field equation gets from the coupling function a term that is linear in $\varphi$, the resulting Klein-Gordan type equation 
\begin{equation}
    \nabla_\mu \nabla^\mu \varphi + \frac{df}{d\varphi} R^2_{\rm GB} =  \nabla_\mu \nabla^\mu \varphi - m^2_{\rm eff} \varphi =0 \ ,
\end{equation}
contains an effective mass term.
This equation is very reminiscent of Eq.~(\ref{phi_mat}) considered in the case of matter-induced spontaneously scalarized neutron stars.
Indeed as realized in \cite{Doneva:2017bvd,Silva:2017uqg,Antoniou:2017acq} this type of coupling function can give rise to curvature-induced spontaneously scalarized black holes, with the matter source term now replaced by the GB term, i.e., a higher curvature term.

We exhibit the domain of existence of static spontaneously scalarized black holes in Fig.~\ref{fig8} for two coupling functions, for exponential coupling $f(\varphi)=\frac{ \lambda^2}{12} \left(1- e^{-6\varphi^2}\right)$ in Fig.~\ref{fig8}(a) and quadratic coupling $f(\varphi)=  \frac{ \lambda^2}{2} \varphi^2$ in Fig.~\ref{fig8}(b).
The symmetry $\varphi \to - \varphi$ of the coupling functions leads to symmetric sets of EsGB solutions. 
When the curvature source term for the scalar field becomes sufficiently strong, a tachyonic instability sets in, and the fundamental branch ($n=0$) of static scalarized black holes bifurcates from the Schwarzschild solution, whose GB term is given by $R^2_{\rm GB} = {48 M^2}/{r^6}$.
With increasing curvature, and thus smaller horizon radii and masses, a countable family of radially excited ($n>0$) EsGB black holes branches off from the Schwarzschild solution.

\begin{figure}[ht]
\begin{center}
\mbox{
\includegraphics[width=4.5cm, angle =270]{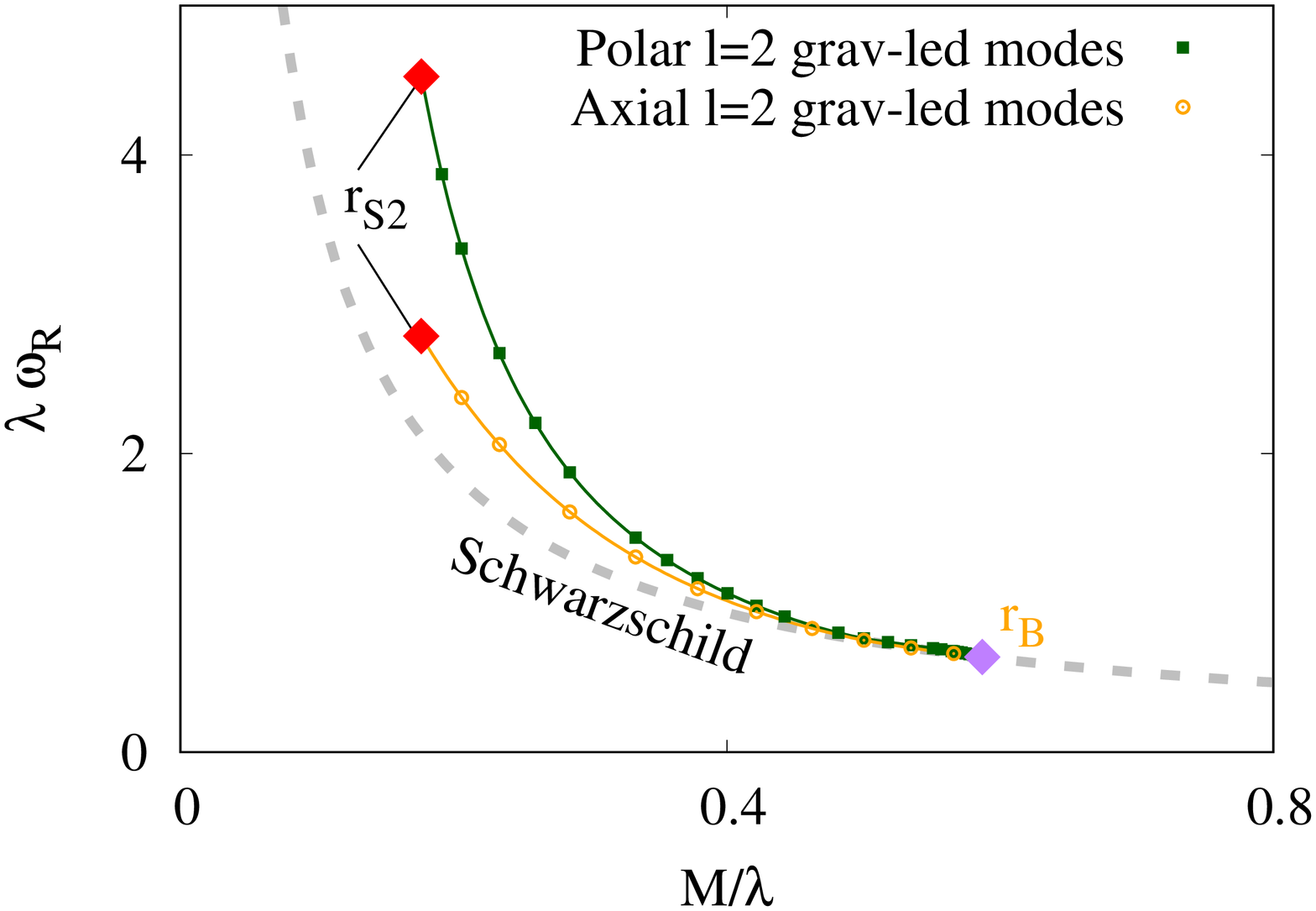}
\includegraphics[width=4.5cm, angle =270]{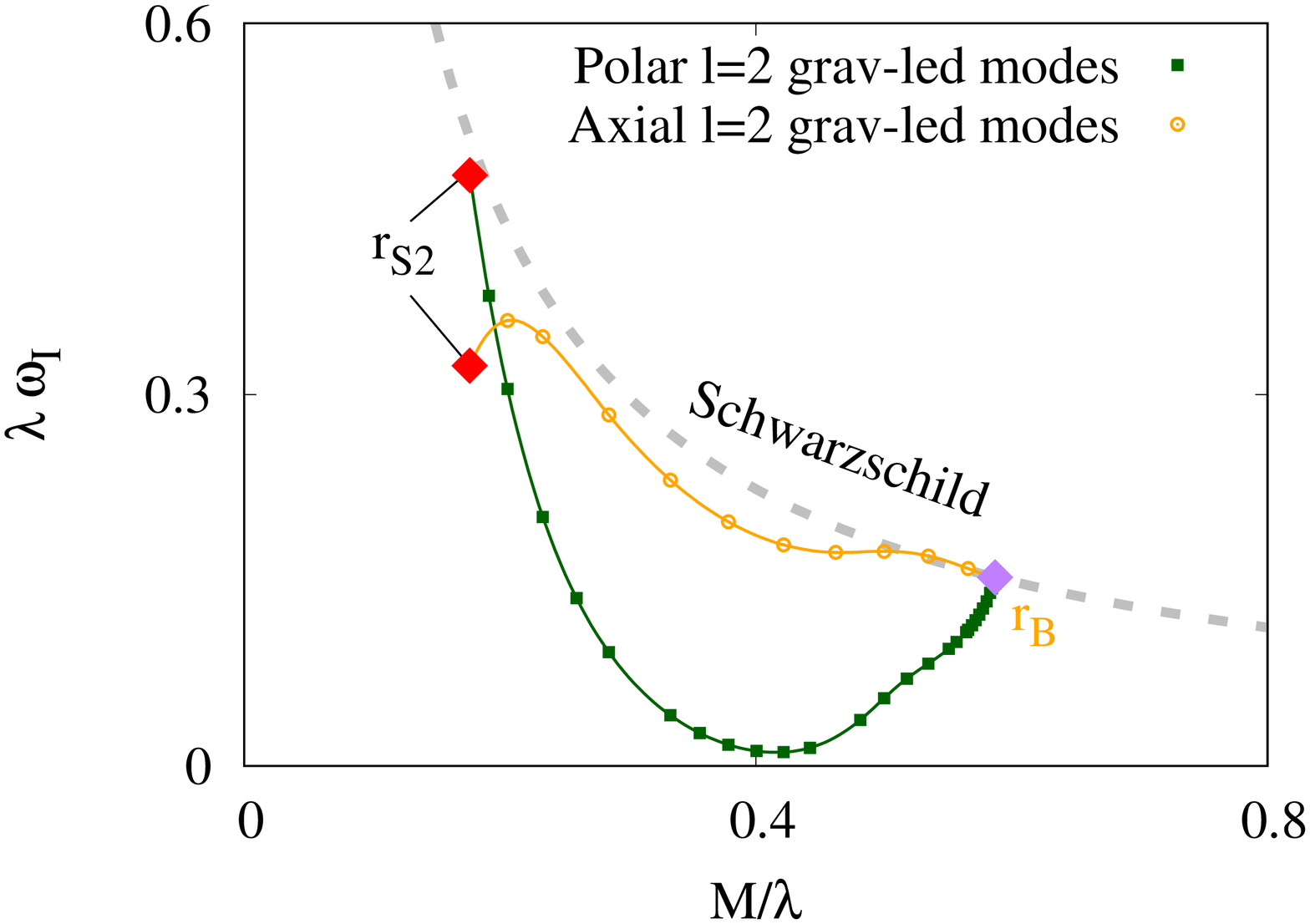}
}
\end{center}
\vspace{-0.5cm}
\caption{{Polar and axial $l=2$ QNMs of static spontaneously scalarized EsGB black holes with exponential coupling: (a) scaled frequency $\lambda \omega_R$ vs scaled mass $M/\lambda$; (b) scaled decay rate $\lambda \omega_I$ vs $M/\lambda$.
Shown are also the Schwarzschild modes, the bifurcation point from Schwarzschild $r_{\rm B}$, and the point $r_{\rm S 2}$, where hyperbolicity is lost for the axial modes \cite{Blazquez-Salcedo:2020rhf,Blazquez-Salcedo:2020caw}.}}
\label{fig9}
\end{figure}

Inspection of the figures shows, that only the fundamental black hole branch of the exponential coupling features a concave set of solutions. 
All other sets of solutions are convex. 
This distinction has serious consequences for the stability of the solutions.
As demonstrated in \cite{Doneva:2017bvd,Blazquez-Salcedo:2018jnn}, only the $n=0$ branch is stable against radial perturbations.
All other branches possess radially unstable modes, that branch off from the unstable Schwarzschild modes.
We recall, that the Schwarzschild solution itself has a zero mode at each bifurcation point, that subsequently turn into unstable modes.

When following the fundamental branch to smaller values of the mass for fixed coupling constant, a point $r_{\rm S 1}$ is reached where the radial perturbation equations lose hyperbolicity, i.e., the employed formalism breaks down.
To investigate, whether the solutions on the fundamental branch are mode stable between the bifurcation point from Schwarzschild $r_{\rm B}$ and $r_{\rm S 1}$, the polar and axial modes should be free of instabilities.
As shown in \cite{Blazquez-Salcedo:2020rhf,Blazquez-Salcedo:2020caw} this is indeed the case, except that the axial perturbation equations lose hyperbolicity slightly earlier at $r_{\rm S 2}$.
The lowest polar and axial $l=2$ grav-led modes are shown in Fig.~\ref{fig9} and compared to the Schwarzschild modes.
For the scalarized black holes isospectrality is clearly broken.
We note, that stability can also be achieved for a quadratic coupling function, when a potential for the scalar is included
\cite{Macedo:2019sem}.

We finally address the rotating spontaneously scalarized EsGB solutions and their properties \cite{Cunha:2019dwb,Collodel:2019kkx,Dima:2020yac,Hod:2020jjy,Doneva:2020nbb,Herdeiro:2020wei,Berti:2020kgk}.
To this end we inspect the GB term of the Kerr solutions
\begin{equation}
    R^2_{\rm GB} = \frac{48 M^2}{(r^2 + \chi^2)^6} 
\left( r^6 - 15 r^4 \chi^2 + 15 r^2 \chi^4 - \chi^6 \right)
\ , \ \ \ \chi=a \cos \theta \ ,
\end{equation}
($a=J/M$), since in the case of rotation the Kerr black holes should be considered in the source term of the scalar equation.
Obviously, the static spontaneously scalarized EsGB black holes can be set into rotation, and thus there are rotating EsGB black holes for positive coupling constant \cite{Cunha:2019dwb,Collodel:2019kkx}.
However, as already noted in the case of neutron stars, there is a second type of scalarized black holes, that arise for negative coupling constant \cite{Dima:2020yac,Hod:2020jjy,Doneva:2020nbb,Herdeiro:2020wei,Berti:2020kgk}.
In this case, the GB invariant should be negative in some regions of space with high curvature, which can be achieved for sufficiently high spins.
In fact, $J/M^2 > 0.5$ is needed for this spin-induced spontaneous scalarization to occur \cite{Dima:2020yac,Hod:2020jjy}.

\begin{figure}[ht]
\begin{center}
\mbox{
\includegraphics[width=6.5cm]{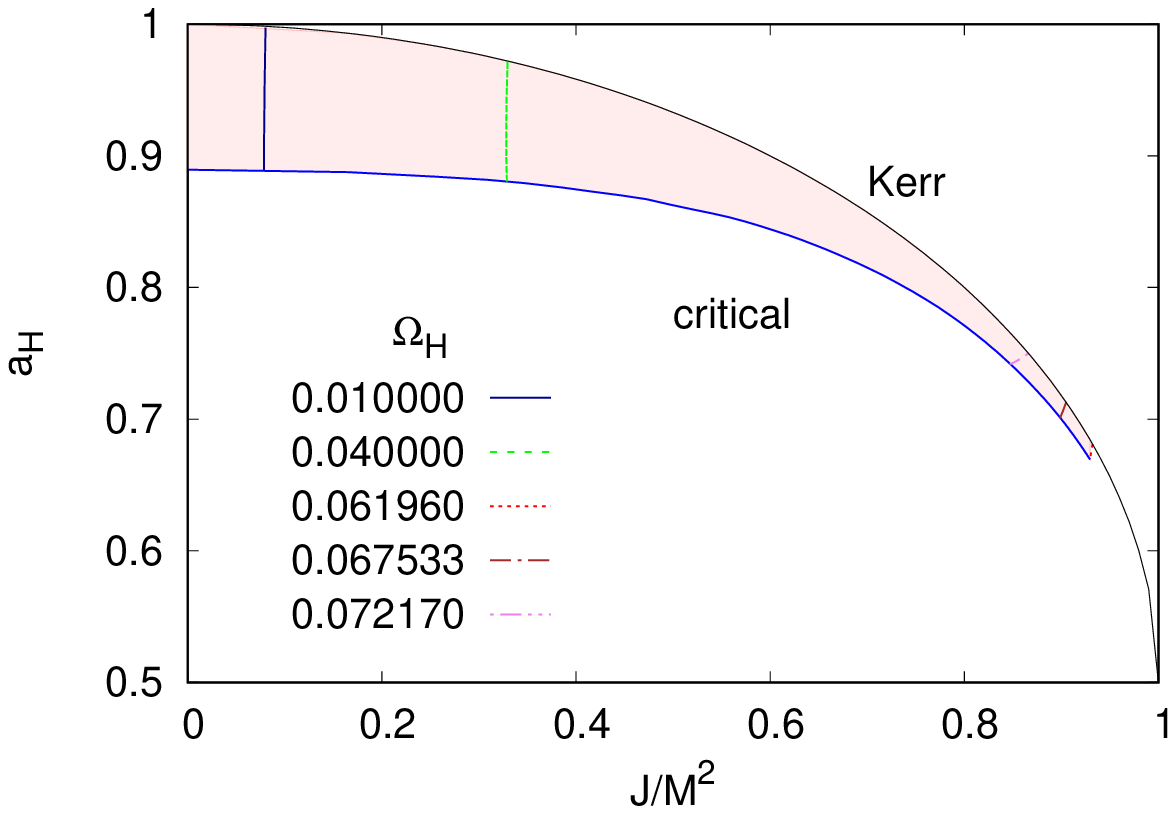}
\includegraphics[width=6.5cm]{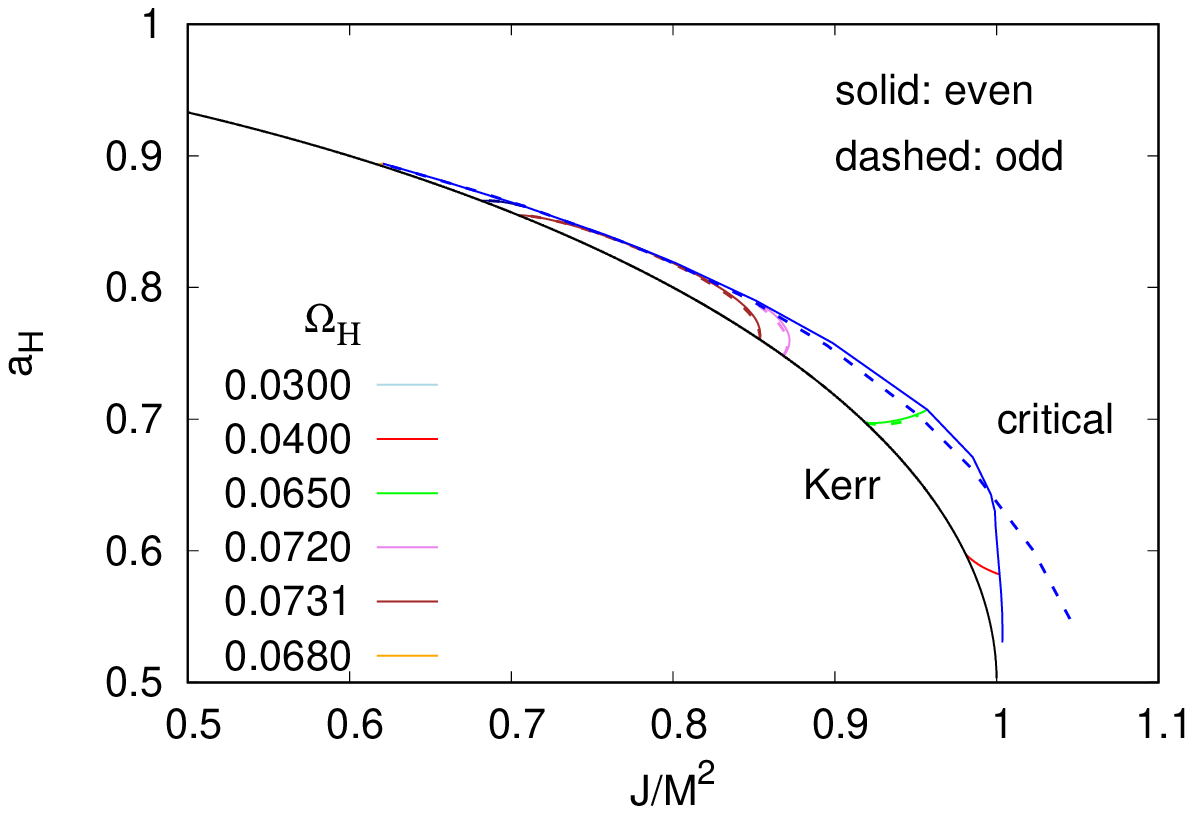}
}
\end{center}
\vspace{-0.5cm}
\caption{{Domain of existence of rotating spontaneously scalarized GB black holes with quadratic coupling function: (a) scaled horizon area $a_H=A_H/16 \pi M^2 $ vs scaled angular momentum $j=J/M^2$ for positive coupling constant; (b) $a_H=A_H/16 \pi M^2 $ vs $j=J/M^2$ for spin-induced black holes for negative coupling constant, both for even and odd scalar field. 
The domains are delimited by Kerr black holes, and critical solutions. {Also shown are curves of fixed horizon angular velocity $\Omega_H$} \cite{Collodel:2019kkx,Berti:2020kgk}.
}}
\label{fig10}
\end{figure}

As expected from these considerations the domain of existence of rotating EsGB black holes with positive coupling decreases, when the spin increases \cite{Cunha:2019dwb,Collodel:2019kkx}.
This effect is most pronounced for the exponential coupling, where the domain of existence of static black holes is much larger than for the case of the quadratic coupling, shown in Fig.~\ref{fig10}(a).
Moreover, these rotating solutions arise largely from stable static solutions.
Thus they are expected to retain stability.
Calculation of their entropy in fact also points to their stability, since their entropy is always larger than the entropy of comparable Kerr black holes \cite{Cunha:2019dwb}.
This is in contrast to the case of the quadratic coupling where the entropy is smaller \cite{Collodel:2019kkx}.
Of course, in this case already the static solutions are unstable.
Since in the exponential case, the horizon area of the scalarized black holes can be considerably smaller than the horizon area of comparable Kerr black holes, this opens the possibility to test their shadow by observations and thus put bounds on the coupling constant \cite{Cunha:2019dwb}.

The domain of existence of spin-induced EsGB black holes has been determined in \cite{Herdeiro:2020wei} for the exponential coupling and \cite{Berti:2020kgk} for the quadratic coupling.
Interestingly, in this case both coupling functions lead to similar results.
In particular, in both cases the horizon area of spin-induced black holes is smaller than the horizon area of comparable Kerr black holes, whereas their entropy is larger.
This indicates that the rapidly spinning EsGB black holes with spin-induced spontaneous scalarization might be stable even in the case of the quadratic coupling.
However, a stability analysis of the rotating EsGB black holes is still missing.

We end this discussion by mentioning that recent studies have applied techniques of numerical relativity to EsGB theories with spontaneous scalarization of black holes.
These studies address, for instance, dynamical scalarization and descalarization in black hole mergers or the dynamical formation of scalarized black hole in the collapse of burned out stars
\cite{Witek:2018dmd,Witek:2020uzz,Silva:2020omi,Kuan:2021lol,Doneva:2021dqn,East:2021bqk}.

\section{Conclusions}

New observations on the ground and in space are giving us unprecedented insight on the Universe.
This concerns both the evolution of the Universe as well as the celestial bodies inside it and their evolution and dynamics.
Many of the associated theoretical studies have focused in recent years on alternative gravities to overcome the limitations and difficulties encountered.
Indeed, alternative gravities have open up many promising directions in the ongoing quest (see e.g.~\cite{Faraoni:2010pgm,Berti:2015itd,CANTATA:2021ktz}).

Here we have addressed neutron stars and black holes in a set of alternative gravities, that involve an additional scalar degree of freedom.
The presence of such a new degree of freedom then allows for a variety of interesting phenomena.
Foremost the spectrum of compact objects is considerably widened, since beside the gravitational quadrupole (and higher mode) radiation, also monopole and dipole radiation is allowed, while the $l\ge 2$  gravitational-led modes are supplemented by scalar-led modes.
Viable alternative gravities must respect the associated bounds set by current and future observations.

Also, depending on the theory, neutron stars can, for instance, undergo matter-induced spontaneous scalarization, that is recurrent in black holes in the form of curvature-induced spontaneous scalarization.
The resulting compact objects may then exhibit genuine properties distinguishing them from their GR brethren.
Neutron stars may then possess deviating \textit{universal relations}, while black holes may exhibit distinct shadows, etc., to be compared with observations, instrumental to obtain bounds.

Of considerable future relevance should be the communication between the groups working on alternative gravities as applied to cosmological issues and to compact objects.
In particular, cosmologically attractive theories should be scrutinized with respect to their predictions for compact objects.
Here first steps have been undertaken, and the future will certainly see more of these promising lines of research, also encouraged by the organizers of \textbf{ALTECOSMOFUN’21} by the schedule of the meeting.

\authorcontributions{All authors contributed substantially.}

\funding{This research was funded in part by DFG RTG 1620 \textit{Models of Gravity} and FCT project PTDC/FIS-AST/3041/2020.
}

\acknowledgments{We would like to thank the organizers for the invitation to the interesting conference \textbf{Alternative Gravities and Fundamental Cosmology - ALTECOSMOFUN'21}.
We would also like to thank our collaborators: Emanuele Berti, Vitor Cardoso, Lucas G. Collodel, Daniela D. Doneva, Valeria Ferrari, Luis M. Gonz\'alez-Romero, Leonardo Gualtieri, Sarah Kahlen, Panagiota Kanti, Fech Scen Khoo, Caio F. B. Macedo, Sindy Mojica, Zahra A. Motahar, Francisco Navarro-L\'errida, Petya Nedkova, Paolo Pani, Vincent Preut, Eugen Radu, Kalin V. Staykov, Stoytcho S. Yazadjiev. 
We gratefully acknowledge support by the DFG Research Training Group 1620 \textit{Models of Gravity} and the COST Actions CA15117 and CA16104. 
JLBS would like to acknowledge support from FCT project PTDC/FIS-AST/3041/2020.}

\conflictsofinterest{The authors declare no conflict of interest.}

\begin{adjustwidth}{-\extralength}{0cm}

\end{adjustwidth}
\end{document}